\documentclass[12pt]{article}
\pdfoutput=1
\usepackage{amsmath,amssymb}
\usepackage{hyperref,hep}
\usepackage{graphicx}
\newcommand{\be}{\begin{equation}}
\newcommand{\ee}{\end{equation}}
\newcommand{\bea}{\begin{eqnarray}}
\newcommand{\eea}{\end{eqnarray}}

\begin{document}

\makeatletter
\renewcommand{\theequation}{\thesection.\arabic{equation}}
\@addtoreset{equation}{section}
\makeatother

\baselineskip 18pt

\begin{titlepage}

\vfill

\begin{flushright}
%tags
\end{flushright}

\vfill

\begin{center}
   {\Large\bf
   Nonlinear conductivity and the ringdown of currents in metallic holography
   }
  \vskip 1.5cm
      Benjamin Withers\\
   \vskip .6cm
	  b.withers@qmul.ac.uk\\
     \vskip .6cm
     \textit{School of Mathematical Sciences, Queen Mary University of London, E1 4NS, UK\\
      \vspace{1em}
      Centre for Research in String Theory, School of Physics and Astronomy, \\Queen Mary University of London, E1 4NS, UK\\
      \vspace{1em}
     DAMTP, University of Cambridge, CB3 0WA, UK
     }
\end{center}

\vfill

\begin{center}
\textbf{Abstract}
\end{center}

\begin{quote}
We study the electric and heat current response resulting from an electric field quench in a holographic model of momentum relaxation at nonzero charge density. After turning the electric field off, currents return to equilibrium as governed by the vector quasi-normal modes of the dual black brane, whose spectrum depends qualitatively on a parameter controlling the strength of inhomogeneity. We explore the dynamical phase diagram as a function of this parameter, showing that signatures of incoherent transport become identifiable as an oscillatory ringdown of the heat current. We also study nonlinear conductivity by holding the electric field constant. For small electric fields a balance is reached between the driving electric field and the momentum sink -- a steady state described by DC linear response. For large electric fields Joule heating becomes important and the black branes exhibit significant time dependence. In a regime where the rate of temperature increase is small, the nonlinear electrical conductivity is well approximated by the DC linear response calculation at an appropriate effective temperature.
\end{quote}

\vfill

\end{titlepage}

\tableofcontents

\section{Introduction}
Transport properties of the strongly coupled metals and insulators of gauge/gravity duality has been the subject of much recent interest. 
At a practical level this is enabled by the proliferation of stationary black brane solutions with AdS asymptotics which incorporate momentum relaxation through explicitly sourced inhomogeneities, either directly in the form of inhomogeneous lattices \cite{Horowitz:2012ky,Horowitz:2012gs,Donos:2012js,Horowitz:2013jaa,Ling:2013nxa,Chesler:2013qla,Balasubramanian:2013yqa,Donos:2014cya,Donos:2014oha,Donos:2014yya,Rangamani:2015hka}, disorder \cite{Davison:2013txa, Lucas:2014zea, Hartnoll:2014cua, Lucas:2015vna, O'Keeffe:2015awa, Hartnoll:2015faa}, or other constructions motivated by their bulk simplicity \cite{Donos:2013eha, Andrade:2013gsa,Donos:2014uba,Taylor:2014tka,Andrade:2016tbr}.
For each equilibrium brane, thermoelectric conductivities can be obtained at finite temperature using a linear response calculation. DC conductivities can be evaluated provided certain data at the black brane horizon is known \cite{Iqbal:2008by, Donos:2014cya}, and in some generality \cite{Donos:2014yya,Donos:2015gia,Banks:2015wha}.

In this paper we turn to an investigation of the nonlinear response of such holographic models, following the quench of an applied electric field.
From the gravitational point of view, this amounts to an investigation of the nonlinear dynamics associated to black branes in AdS with inhomogeneous Dirichlet boundary conditions.
One should expect qualitatively different dynamical behaviour for bulk evolutions with such boundary conditions, as compared with those with homogeneous boundary conditions. At the very least the boundary CFT dynamics are distinct, with the former conserving momentum either only approximately or not at all.\footnote{Previous work on the dynamics associated to momentum relaxation was discussed in a complementary class of holographic experiments in \cite{Balasubramanian:2013yqa}.}\footnote{See \cite{Davison:2013jba, Blake:2015epa} for approaches to a hydrodynamic-like description when momentum relaxation is weak, in the context of axion models.} Thus at late times we do not expect to find hydrodynamic behaviour, but rather only exponentially fast decay described by the quasi-normal modes (QNMs) associated to momentum relaxation. 
The QNM spectrum itself is known to change qualitatively with the strength of inhomogeneity, resulting in transitions between coherent and incoherent metals \cite{Davison:2014lua,Davison:2015bea,Andrade:2015hpa}, offering up further interesting features for the bulk dynamics.

From the field theory point of view, an electric field is a natural tool to quench an inhomogeneous system with finite charge density.\footnote{For studies of electric field quenching in a probe brane context see for example \cite{Hashimoto:2014yza, Ali-Akbari:2015gba, Ali-Akbari:2015hoa}.} The non-conservation of momentum allows the possibility of approaching a finite steady state, at least in the linear response regime. 
In particular, as a special case this includes the relaxation of currents to zero in the laboratory frame after the electric field is turned off. Here we can study the imprint of the QNM spectrum on the current relaxation, in particular how it is affected by the aforementioned coherent to incoherent shifts in the QNM spectrum.\footnote{We thank Richard Davison for encouraging pursuit of these features at nonzero charge density.}

For constant electric fields we compute nonlinear electric and thermoelectric conductivity. Here the ideal scenario would involve a steady state solution, perhaps utilising an external heat bath. This is precisely the kind of scenario that is argued to arise in some probe brane constructions \cite{Karch:2007pd, Albash:2007bq, Karch:2010kt, Sonner:2012if, Berridge:2013yba, Baggioli:2016oju, Kundu:2013eba,Kundu:2015qda}. In the present backreacted context a steady state will not be reached since energy is continually added and the system exhibits Joule heating without bound. It would be interesting to consider models which realise a coupling to an external heat bath incorporating backreaction -- we consider some of results presented here as a precursor to this problem, which we shall leave to future work. 

Remarkably however, even in the absence of a steady state, nonlinear electrical conductivity has been computed exactly in the absence of charge \cite{Horowitz:2013mia}, explored using a Vaidya spacetime.
Similarly in this paper we will be able to make clean statements about the nonlinear electric and thermoelectric conductivities at finite charge density in the absence of a steady state, at least in a limit where the rate of heating is low. 

To motivate our choice of bulk model, we note that at finite charge density, the electric field will drive both electric and heat currents. It is therefore desirable that the model used incorporates some form of momentum relaxation.
The minimal ingredients required are provided by an Einstein-Maxwell-axion model \cite{Andrade:2013gsa}. Each axion field, $\phi_I$, is dual to an operator $O_I$ sourced by the value of $\phi_I$ at the AdS boundary, $\phi^{(0)}_I$. To break translational invariance the sources are taken to be linear in the boundary spatial coordinates $\phi^{(0)}_I\propto x^I$. If two axions are used, the equilibrium state can be made isotropic, although driving the system with an electric field will break isotropy in general. The axion model brings a clear computational benefit since the bulk geometry remains independent of $x^I$.\footnote{The only $x^I$ dependence appears in $\phi_I$ and the gauge field, $A$, where it is known in advance.}

Some basic aspects of the role of the electric field in the axion model can be understood in reference to the Ward identities,
\bea
\partial_\mu \left<T^\mu_{~~\nu}\right> &=& \partial_\nu \phi^{(0)}_I \left<O_I\right> + (F^{(0)})_{\nu}^{~~\mu} \left<J_\mu\right> \label{wardT}\\
\partial_\mu \left<J^\mu \right> &=& 0 \label{wardJ}
\eea
where $T_{\mu\nu}$ is the QFT stress tensor, $J_\mu$ the U(1) current and $F^{(0)} = dA^{(0)}$ is a classical, boundary field strength. Throughout this paper the sources are taken to be,
\be
\phi^{(0)}_I = k x^I , \qquad F^{(0)} = E(t) dx\wedge dt
\ee
where $I=1,2$ labels the two spatial boundary directions, $x = x^1, y = x^2$. With the exception of $\phi^{(0)}_I$ and $A^{(0)}$ all quantities here are independent of $x^I$. Inserting these expressions into \eqref{wardT} we find,
\bea
\partial_t \epsilon &=& EJ \label{joule}\\
\partial_t J_E &=& k \left<O_1\right> + E \rho\label{balance} 
\eea
where $\epsilon= \left<T^{tt}\right>$ is the energy density, $\rho = \left<J^t\right>$ is the charge density, $J = \left<J^x\right>$ is the electrical current, $J_E = \left<T^{tx}\right>$ is the energy current.\footnote{The $y$-component of \eqref{wardT} is trivially satisfied for the solutions here, since the electric field is taken in the $x$-direction and $O_2$ does not acquire a nonzero vev.} The first equation \eqref{joule} shows that energy is injected by the electric field, which results in Joule heating. The second, \eqref{balance}, accounts for the momentum, and because of the axion sink term there is the possibility of a steady state where the terms on the hand side cancel, balancing momentum relaxation with the driving electric field. Finally we note that \eqref{wardJ} gives $\partial_t \rho = 0$.

Since we shall be subject to the nonlinearities of the full evolution of the Einstein-Maxwell-axion system, we resort to a numerical integration of the bulk equations of motion. Details of the numerics are given in appendices \ref{numerics} and \ref{convergenceconvergenceconvergence}. We will study two different cases for the profiles of $E(t)$:
\begin{itemize}
\item \textbf{Top hat $E(t)$ -- current relaxation}. In the first case we smoothly turn on $E(t)$ hold it at some constant value, and then turn it off again. This experiment will confirm crucial aspects of the model, first and foremost, its stability. We confirm that the system returns to a member of the family of equilibrium solutions at a rate consistent with its longest lived QNMs.  In particular we will examine the QNM spectrum for different values of the momentum relaxation parameter $k$, and show its imprint on current relaxation. We will see a qualitative change in the relaxation of the heat current following a quench to an incoherent regime. Additionally, the return to equilibrium will allow us to get a complete picture of the gravitational situation, allowing the identification of the black hole event horizon.

\item \textbf{Step $E(t)$ -- nonlinear conductivity}. For the second case, we smoothly turn on $E(t)$ and hold it at a constant value $E_f$.
At sufficiently late times, transient behaviour associated to the quenched electric field dies down and we enter a regime where the system is effectively responding only to a constant $E = E_f$. When $E$ is constant we choose to characterise the response of the system using electrical ($\sigma$) and thermoelectric ($\bar{\alpha}$) DC conductivities, defined as follows,
\bea
J &=& \sigma E_f,\\
Q \equiv J_E-\mu J &=& \bar{\alpha} T\; E_f
\eea
where $Q$ is the heat current in the absence of thermal gradients \cite{Donos:2014cya}. Here we have restricted to the $xx$-entries in each conductivity matrix, with the $yx$-entries vanishing. The thermal conductivity will not play a role since there is no temperature gradient. Note that $\sigma$ and $\bar{\alpha}$ will be time dependent in general due to the effect of heating, and they may also depend explicitly on $E_f$.
Such examples of perpetually driven CFTs are less well studied in favour of a system which eventually returns to equilibrium. An interesting set of examples are provided in \cite{Rangamani:2015sha}. 
\end{itemize}
In the context of the second class of experiments we can gain analytic control over $\sigma,\bar{\alpha}$ in two limits of parameter space. For sufficiently small $E_f$, linear response can be used, finding \cite{Andrade:2013gsa, Donos:2014cya}
\be
\sigma = 1 + \frac{\mu^2}{k^2}, \qquad \bar{\alpha} = \frac{4\pi \rho}{k^2}. \label{linearcond}
\ee
In this regime $\left<O_1\right> = -E_f \rho/k$ and the right hand sides of \eqref{joule} and \eqref{balance} vanish at linear order in $E_f$, corresponding to a steady state. For $\rho=0$ one can go beyond linear response analytically for any $E_f$, finding
\be
\sigma = 1, \qquad \bar{\alpha} = 0. \label{neutralcond}
\ee
In fact, these expressions apply for any choice $E(t)$, showing that $J(t)$ responds instantaneously to $E(t)$. These conductivities are extracted from a Vaidya-like spacetime which can be constructed with axions \cite{Bardoux:2012aw}. This generalises the conductivity results of \cite{Karch:2010kt, Horowitz:2013mia} to include a momentum relaxation parameter.\footnote{There is a difference of convention of electric charge to \cite{Horowitz:2013mia}, as can be seen by examining the action, which accounts for the factor of $4$ difference.} In the absence of net charge, the added feature of momentum relaxation may be somewhat redundant, nevertheless these solutions serve as a useful reference point. Indeed we will show that an instantaneous electric current is the first response of the system in the charged examples that follow. 

Away from these two limits the responses of $J$ and $Q$ are subject to the nonlinearities of the system. Remarkably, we find that the expression for $\sigma$ in \eqref{linearcond} provides an excellent accounting of the evolution of $J$ far from linear response, provided some local notion of temperature during the evolution. This applies even when Joule heating introduces significant time dependence of the black brane. On the other hand $\bar{\alpha}$ does not appear amenable to the same treatment, exhibiting nonlinear dependence on $E_f$.

The paper is organised as follows. In section \ref{model}, we present the equilibrium Einstein-Maxwell-axion black branes. In section \ref{neutral} we review Vaidya-like solutions to the Einstein-axion system, and for the case of an electric field compute the nonlinear $\sigma$ and $\bar{\alpha}$ at $\rho=0$, as given in \eqref{neutralcond}. We then turn to the numerical solutions for a top hat electric field profile in section \ref{hatE} including a discussion of QNMs and the transition to an incoherent regime.  We examine nonlinear conductivity in section \ref{constE}. We conclude in section \ref{comments}. In appendix \ref{numerics} we detail the numerical setup used, with convergence tests presented in Appendix \ref{convergenceconvergenceconvergence}.

\section{Gravitational model and equilibrium black branes\label{model}}
Let us briefly recap the model and present the black brane solutions of \cite{Bardoux:2012aw,Andrade:2013gsa}, presented in ingoing Eddington-Finkelstein coordinates and in a generalised coordinate frame on the boundary. These give the equilibrium, finite temperature, charged metallic state which we will later force out of equilibrium using $E(t)$. We employ the action specialised to $D=4$,
\begin{equation}
	S = \frac{1}{2\kappa_4^2}\int_M \sqrt{-g} \left[ R - 2 \Lambda - \frac{1}{2} \sum_{I=1}^{2} (\partial \phi_I)^2  - \frac{1}{4} F^2 \right ] d^{4} x\label{the model}
\end{equation}
where $\Lambda = - 3\ell^{-2}$ and $F=dA$. The units employed throughout set $2\kappa_4^2=1$ and we set $\ell=1$.\footnote{As usual we also supplement $S$ with appropriate Gibbons-Hawking terms and counterterms for holographic renormalisation.} This theory admits a family of equilibrium black branes which can be written in the following form,
\bea
ds^2 &=& \frac{1}{r^2}\left(-f(r) (u_\mu dx^\mu)^2 + 2u_\mu dx^\mu  dr  +  (\eta_{\mu\nu} + u_\mu u_\nu)dx^\mu dx^\nu\right)\\
A &=& h(r) u_\mu dx^\mu,\quad \phi_I = k n^I_\mu x^\mu
\eea
where $x^\mu = (v,x,y)$ with constant 3-vectors $u$ and $n^I$, $I=1,2$. The functions appearing are given as follows,
\be
f(r) = 1 - \frac{1}{2}k^2r^2 -m r^3 +\frac{1}{4}\rho^2r^4,\qquad h(r) = \mu - \rho\, r
\ee
which is a solution to the equations of motion provided we satisfy the following orthogonality condition,
\be
\delta_{IJ} n^I_\mu n^J_\nu = \eta_{\mu\nu} + u_\mu u_\nu
\ee
which is equivalent to the 6 relations, $u^2 = -1$, $\eta_{\mu\nu} n^\mu_I n^\nu_J  = \delta_{IJ}$, $n^I_\mu u^\mu = 0$. The stress tensor and current are given by, 
\bea
\left<T^{\mu\nu}\right> &=& \epsilon u^\mu u^\nu + \frac{\epsilon}{2} (\eta^{\mu\nu} + u^{\mu} u^{\nu})\\
\left<J^\mu\right> &=& \rho u^\mu
\eea
where $\epsilon = 2m$.
For convenience we work in a fixed laboratory reference frame, setting $u = \partial_v$ and $n_I = \partial_{x^I}$, which satisfies all of the above conditions for a solution.

The thermodynamical quantities can be written most succinctly in a form parameterised using the event horizon coordinate location $r_0$ (where $f=0$),
\bea
T &=& \frac{1}{4 \pi r_0} \left(3 - \frac{k^2 r_0^2}{2}  -  \frac{\mu^2r_0^2}{4}   \right), \qquad s = \frac{4 \pi}{r_0^{2}}\\
\epsilon = 2 m &=& \frac{2}{r_0^{3}} \left(1- \frac{k^2r_0^2}{2} + \frac{\mu^2 r_0^2}{4}\right)\label{thermoEp}\\
\rho &=& \frac{\mu}{r_0}\label{thermoRho}
\eea
Linear response DC electric and thermoelectric conductivities for this background are given by \eqref{linearcond}. Further results for the transport properties of this model are discussed in \cite{Andrade:2013gsa,Davison:2014lua,Donos:2014cya,Davison:2015bea,Andrade:2015hpa,Hartnoll:2016tri}.

\section{Vaidya-like solutions with axions\label{neutral}}
It is also possible to construct Vaidya-like spacetimes in the presence of the axion sources \cite{Bardoux:2012aw}. Here we review these solutions and compute their nonlinear electric and thermoelectric conductivities in the vein of \cite{Horowitz:2013mia}. Specifically, in the $D=4$ Einstein-axion system, we can add an additional stress tensor source, $\tilde{T}_{ab}$, leading to the equations of motion
\bea
R_{ab} + 3g_{ab} - \frac{1}{2}\sum_I^{2}\partial_ a \phi_I \partial_b \phi_I &=& \kappa^2_4\left(\tilde{T}_{ab} - \frac{1}{2}g_{ab} \tilde{T}\right) \\
\Box\phi_I &=& 0.
\eea
For a null stress tensor of the form, $\tilde{T}_{vv} = r^2 S(v)$ with other entries zero, the following electrically neutral solution is obtained, 
\bea
ds^2 &=& \frac{1}{r^2}\left(-\left(1 -\frac{1}{2}k^2r^2 -m(v) r^3\right) dv^2 - 2dv  dr  +  dx^2+dy^2\right)\\
\phi_I &=& k x^I
\eea
with the sourced energy conservation equation, 
\be
\partial_v (2m(v)) = 2\kappa_4^2 S(v).
\ee
As pointed out in \cite{Bardoux:2012aw} an electric field can be used to generate a stress tensor of this type. Specifically for a Maxwell field in the bulk, $F =  E(v) dx\wedge dv$ gives rise to a stress tensor of the required form, with $S(v) = E(v)^2$. For Einstein-Maxwell in the absence of axions such solutions were  given in \cite{Karch:2010kt,Horowitz:2013mia}.  For this solution the electric and heat currents are $J = E$ and $Q = 0$ -- specialising to a constant electric field $E(v) = E_f$ we see that the nonlinear electric and thermoelectric conductivities are given by \eqref{neutralcond}. Note that this does not straightforwardly extend to $D\neq 4$; the conductivity is no longer dimensionless and is influenced by heating, as was explored in the Einstein-Maxwell context in \cite{Horowitz:2013mia}. Similarly we have not been able to extend these results to $\rho\neq 0$; momentum is nonzero and the solutions do not fit into the form above. The remainder of this paper is focussed on a numerical study of the $\rho \neq 0$ case. The analytical results presented here have been used as a test case for the numerical evolution described in the remainder of this paper.

Finally we note that the time dependent solution in this section can be used to quench the system from an initial equilibrium state with $\epsilon<0$ to a final equilibrium state with exactly $\epsilon =0$, which occurs at $k = \sqrt{2}/r_0$. Despite the remarkable behaviour of the perturbations of this state \cite{Davison:2014lua,Andrade:2015hpa} there appears to be nothing remarkable about a quench so designed.

\section{Current relaxation\label{hatE}}
In this section we drive the system out of equilibrium and then let the energy and electrical currents return to their equilibrium value -- zero, in the laboratory frame defined by the axion sources. We take the electric field profile
\be
E(t) =  \left(\Theta(t) - \Theta(t-t_\ast)\right) E_{c} \label{profilehat}
\ee
where $\Theta(t) \equiv \frac{1}{2}\left(\tanh{\left(\frac{t}{w}\right)}+1\right)$
where $w$ is taken to be some short timescale compared to the intrinsic dynamical response time of the system. 

The return to equilibrium of the currents is governed by the QNM spectrum in the vector sector of perturbations, which changes qualitatively with varying $k/\sqrt{\rho}$ and $T/\sqrt{\rho}$.

\subsection{Vector QNMs at $\rho \neq 0$\label{secQNM}}
One universal statement is that for sufficiently small values of $k^2$ there is a long lived QNM with a purely imaginary frequency, well separated from the other QNMs. In this regime the metal is coherent, with a Drude-peak appearing in the conductivity at low frequency. One can then associate a momentum relaxation rate $\Gamma_\text{rel}$ with the decay rate of this QNM, which has been computed using a matching calculation \cite{Davison:2013jba},
\be
\Gamma_\text{rel} = \tau_{\text{rel}}^{-1} = \frac{sk^2}{6\pi \epsilon}.\label{smallalphatimescale}
\ee
For larger values of $k^2$ one must examine the QNM spectrum in detail.

The vector QNMs are given by ingoing, normalisable solutions to a pair of decoupled equations for gauge invariant master fields, labelled by $(\pm)$, for details see \cite{Andrade:2013gsa, Davison:2015bea}.  At $\rho=0$ the QNM spectrum indicates both coherent and incoherent behaviour is possible \cite{Davison:2014lua,Davison:2015bea,Andrade:2015hpa}. In \cite{Davison:2014lua} the transition between the two behaviours was demonstrated to coincide roughly with the location of a pole collision, producing a damped-oscillatory pole as a dominant contribution to $\left<QQ\right>$. We find that this transition persists at $\rho\neq 0$ in the $(-)$-sector of QNMs, whilst the $(+)$-sector is dominated by a longer lived purely decaying mode. Since in the general case two-point functions of $J$ and $Q$ receive contributions from both $(\pm)$-sectors of QNMs, the relaxation of $J$ and $Q$ are dominated by the longer-lived $(+)$-sector mode.

However, the currents can be decoupled -- see \cite{Davison:2015bea} for details.  To each of the $(\pm)$ sectors one can assign currents, $J_\pm\equiv a_\pm(J_E + \gamma_\pm k J)$, which satisfy $\left<J_\pm J_\mp\right>=0$, where $a_\pm$ is an overall normalisation and 
\be
\gamma_\pm = - \frac{3 \epsilon}{4k \rho} \left(1\pm \sqrt{1+\frac{16 k^2\rho^2}{9\epsilon^2}}\right).\label{gammadef}
\ee
This relation allows one to relate the black hole QNM spectrum in each sector to the thermoelectric response. However, as can be seen in the structure of \eqref{gammadef}, the precise combination of physical currents which achieves this decoupling depends on the details of this particular holographic model.
Universal decoupling is achieved at $\rho=0$ where simply $J_+ = J$ and $J_- = J_E = Q$. But, there is a second limit, $k\to\infty$ where remarkably $J_+ = J$ and $J_- = Q$ \cite{Davison:2015bea}. In other words, if we take large enough $k$, the amplitude of the $(-)$-sector QNMs become parametrically enhanced over the $(+)$-sector QNMs in the phenomenology of $Q$, whilst the $(+)$-sector QNMs remain dominant for $J$.

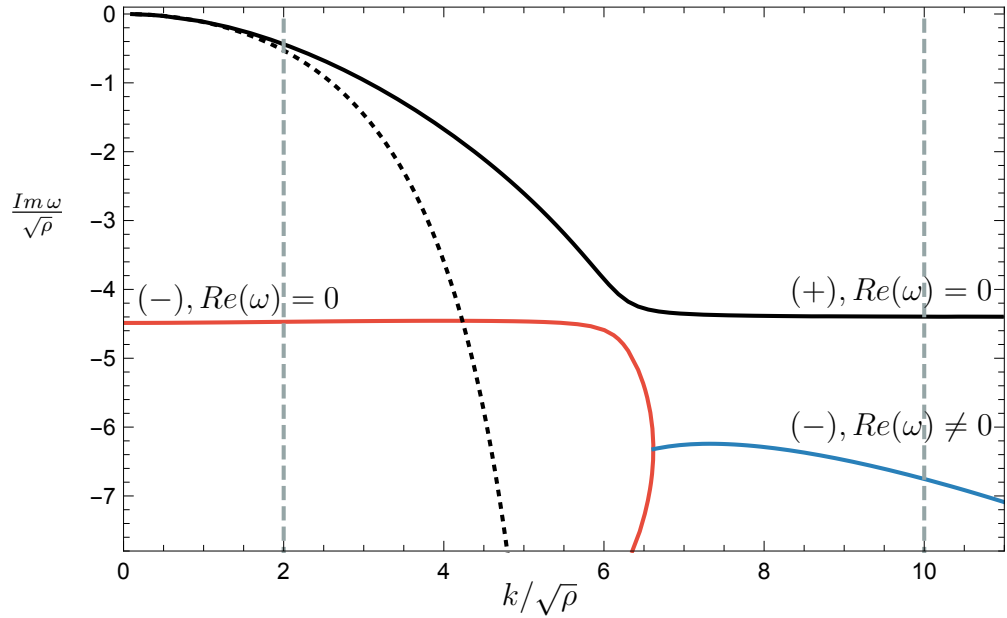
\begin{figure}[h!]
\begin{center}
\includegraphics[width=0.8\textwidth]{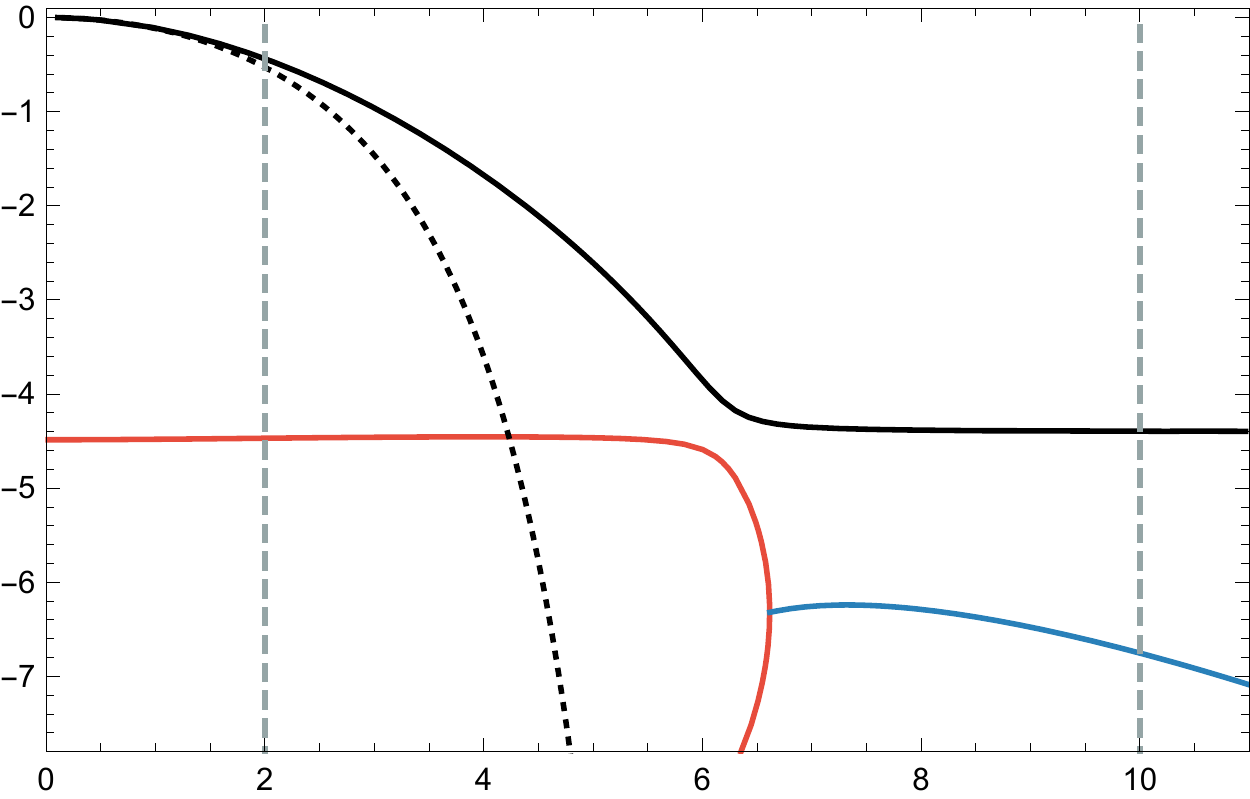}
\begin{picture}(0.1,0.1)(0,0)
\put(-294,108){\makebox(0,0){$(-), Re(\omega) = 0$}}
\put(-47,60){\makebox(0,0){$(-), Re(\omega) \neq 0$}}
\put(-47,111){\makebox(0,0){$(+), Re(\omega) = 0$}}
\put(-370,140){\makebox(0,0){$\frac{Im\,\omega}{\sqrt{\rho}}$}}
\put(-180,-5){\makebox(0,0){$k/\sqrt{\rho}$}}
\end{picture}
\vspace{1em}
\caption{A subset of vector QNMs, including dominant QNMs in each $(\pm)$-sector, at $T/\sqrt{\rho} = 7/10$. The solid black curve is the $(+)$-sector of perturbations, these have $Re(\omega)=0$. The red curve shows $Re(\omega)=0$ modes in the $(-)$-sector of perturbations, whilst the blue curve gives off-axis modes in the $(-)$-sector-- these curves are joined by a pole collision.  The black dashed line is the analytic small $k$ approximation \eqref{smallalphatimescale}.  
The QNMs detailed here govern the relaxation of currents to equilibrium (or to linear steady states). In particular at large $k/\sqrt{\rho}$ the $(-)$-sector modes govern $Q$, whilst the $(+)$-sector modes govern $J$, meaning that the structural changes in the $(-)$-sector spectrum become observable in the relaxation of $Q$.
This is shown using explicit quench examples in section \ref{currquench} at the parameters labelled by the dashed grey lines.
\label{QNMs}}
\end{center}
\end{figure}

We have not performed an exhaustive analysis of this vector QNM spectrum. In figure \ref{QNMs} we show the longest lived QNMs in each sector $(\pm)$ as a function of $k/\sqrt{\rho}$ at $T/\sqrt{\rho} = 7/10$.  We reiterate that even though the $(-)$-sector modes are shorter lived, they can still contribute significantly to the response of $Q$ provided $k$ is large enough. From figure \ref{QNMs} we see that the $(+)$-sector appears to show no significant features, but the $(-)$-sector becomes dominated by oscillatory-decaying modes at large enough $k/\sqrt{\rho}$, just as in the case $\rho =0$. Fortunately, this behaviour is also the regime where the $(-)$-sector can contribute to the response of $Q$. Thus the relaxation of $Q$ will undergo a qualitative change as $k/\sqrt{\rho}$ is dialled.

At this temperature one can see this oscillatory mode results from a pole collision, which closely resembles the neutral case. As one takes higher $T/\sqrt{\rho}$ the case $\rho = 0$ is approached an the red and black curves in figure \ref{QNMs} connect in the vicinity of their closest approach, giving the behaviour observed in \cite{Davison:2014lua,Andrade:2015hpa} where the Drude pole continuously connects to the pole collision. At lower $T/\sqrt{\rho}$ the oscillatory mode remains but the dominant $Re(\omega) =0$ piece of the $(-)$-sector disappears, along with the pole collision.

In the next subsection we examine the far from equilibrium dynamics of $J$ and $Q$ for quenches which return to equilibrium in the two qualitatively different regimes of the QNM spectrum shown in figure \ref{QNMs}.
 
\subsection{Quenches\label{currquench}}
In this section we study electric field quenches which eventually settle down to equilibrium states whose QNMs are given by the spectrum shown in figure \ref{QNMs}. As argued, by dialling $k/\sqrt{\rho}$ we expect to see a qualitative change in the relaxation of $Q$, due to the pole collision and off-axis mode at large $k/\sqrt{\rho}$. The parameters considered here are labelled by the dashed grey lines in figure \ref{QNMs}.

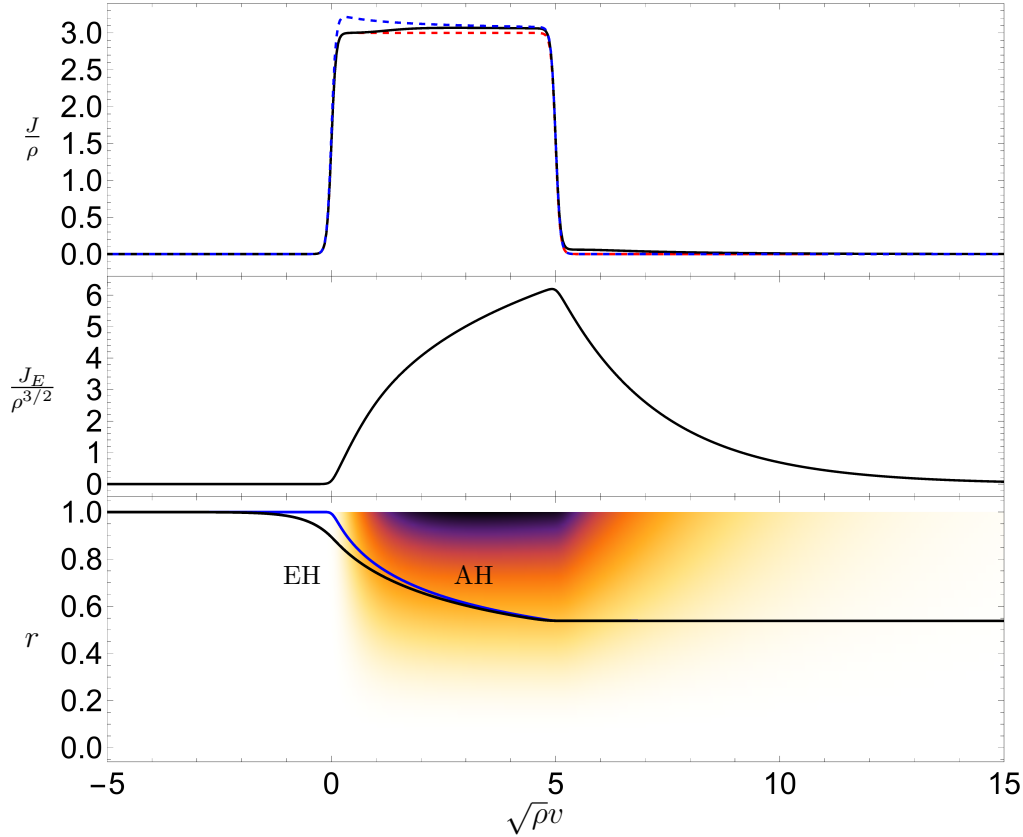
\begin{figure}[h!]
\begin{center}
\includegraphics[width=0.85\textwidth]{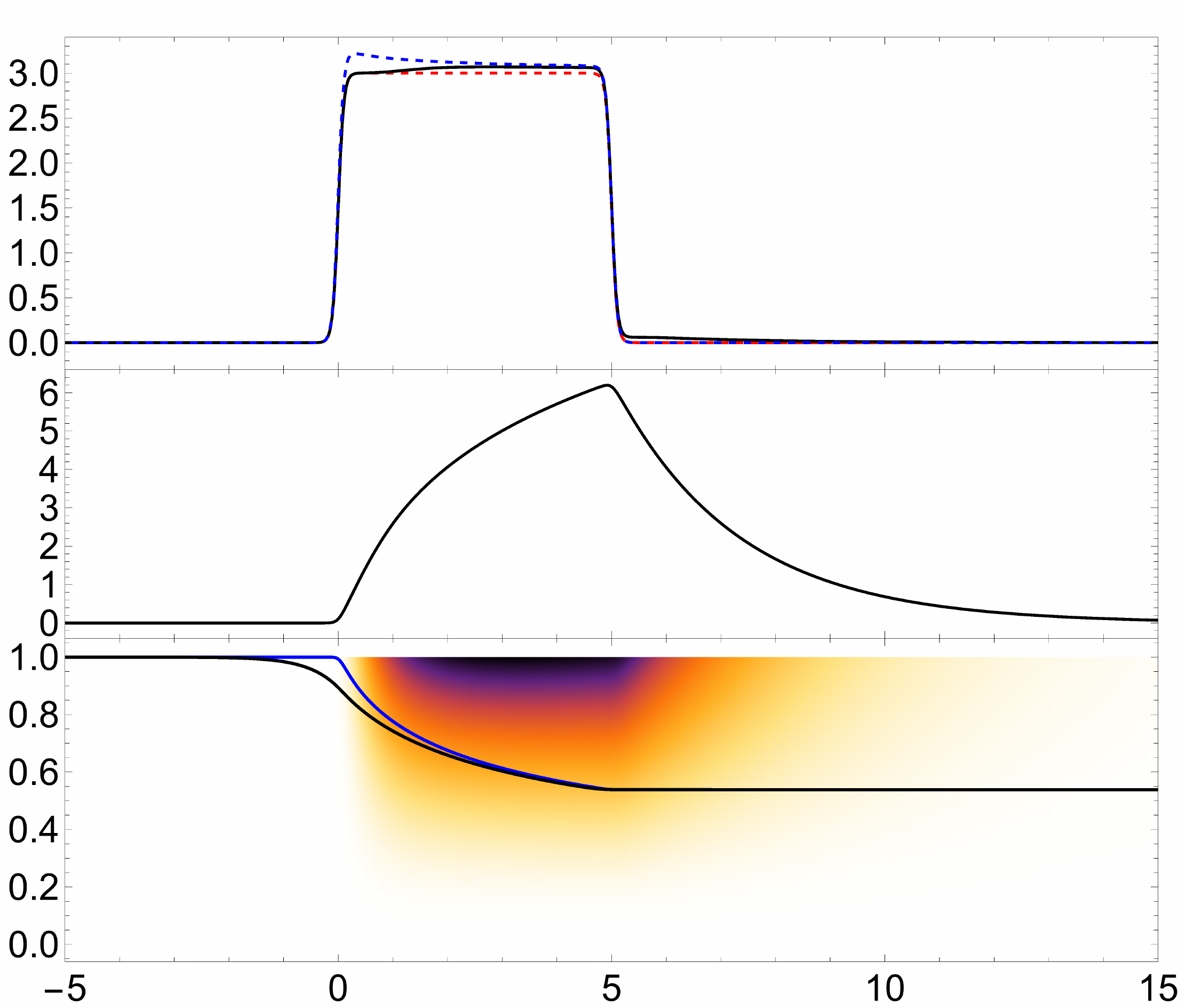}
\begin{picture}(0.1,0.1)(0,0)
\put(-380,250){\makebox(0,0){$\frac{J}{\rho}$}}
\put(-380,155){\makebox(0,0){$\frac{J_E}{\rho^{3/2}}$}}
\put(-380,60){\makebox(0,0){$r$}}
\put(-190,-7){\makebox(0,0){$\sqrt{\rho}v$}}
\put(-213,85){\makebox(0,0){\footnotesize AH}}
\put(-278,85){\makebox(0,0){\footnotesize EH}}
\end{picture}
\vspace{1em}
\caption{Evolution for the top hat electric field profile \eqref{profilehat}. \emph{Top panel:} The electric current. The red dashed curve gives $E(t)/\rho$. The blue dashed curve gives the approximation to $\sigma$ discussed later in section \ref{notsteady}.
\emph{Middle panel:} Energy current. 
\emph{Bottom panel:} Black brane event horizon (EH, black) together with the apparent horizon for the evolution described in appendix \ref{numerics} (AH, blue). The colour illustrates the bulk distribution of the axion field with the linear $x$-dependence subtracted, $\phi_1 - k x$. \label{stepbulk}}
\end{center}
\end{figure}

We first examine the general features of a quench at $k/\sqrt{\rho} = 2$. For this example we choose $E_c/\rho = 3$ with $\sqrt{\rho}t_\ast \simeq 5$, with initial temperature $T_i/\sqrt{\rho} = 3/10$.
In figure \ref{stepbulk} we show the behaviour of $J, J_E$ and the black brane horizon. During the quench in which $E$ is turned on, and shortly after, $J(t) \simeq E(t)$ as in the neutral theory. After $E$ returns to zero the system returns to an equilibrium black brane with $T/\sqrt{\rho} \simeq 0.690$. 

\begin{figure}[h!]
\begin{center}
\vspace{1em}
\includegraphics[width=1\textwidth]{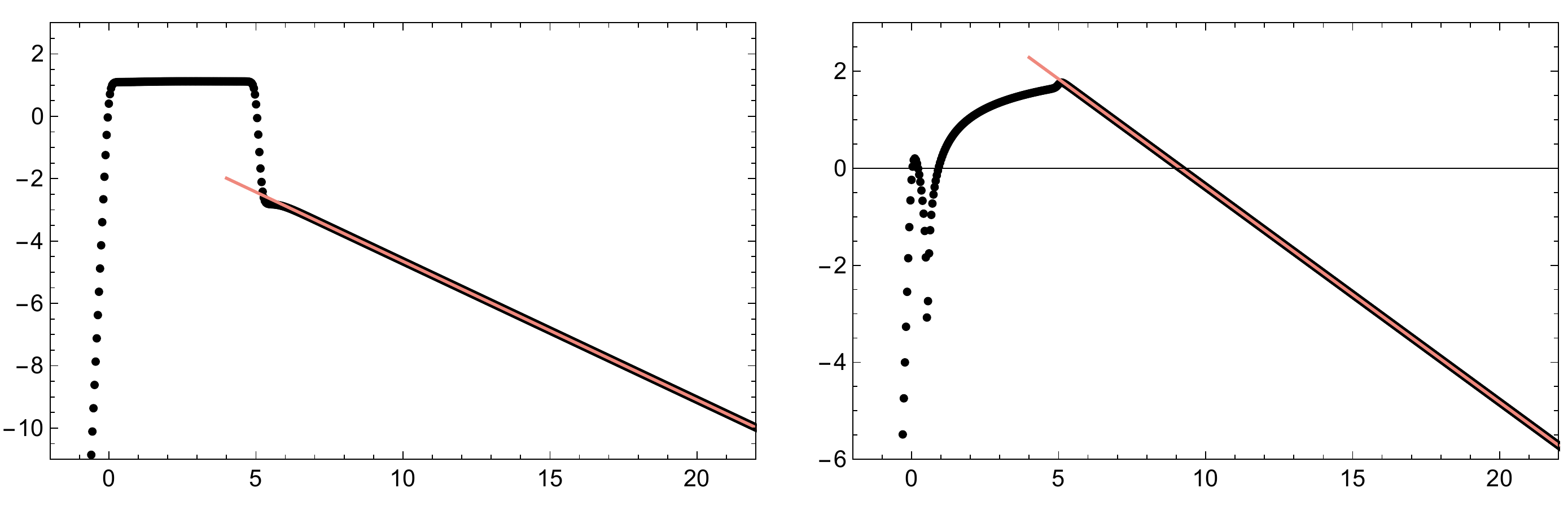}
\begin{picture}(0.1,0.1)(0,0)
\put(-180,163){\makebox(0,0){$\log|J/\rho|$}}
\put(50,163){\makebox(0,0){$\log|Q/\rho^{3/2}|$}}
\put(-98,65){\makebox(0,0){$(+)$}}
\put(145,65){\makebox(0,0){$(+)$}}
\put(-108,15){\makebox(0,0){$\sqrt{\rho}t$}}
\put(116,15){\makebox(0,0){$\sqrt{\rho}t$}}
\put(-108,15){\makebox(0,0){$\sqrt{\rho}t$}}
\end{picture}
\caption{Exponential decay of $J$ (\emph{left panel}) and $Q$ (\emph{right panel}) at $k/\sqrt{\rho}=2$ following a quench to $T/\sqrt{\rho} \simeq 0.690$. The QNM portrait near this temperature is shown in figure \ref{QNMs}. The longest lived QNM belongs to the $(+)$-sector and is purely decaying, dominating both $J$ and $Q$ at this value of $k$. The red curves show a fit to this QNM. \label{ringdown2}}
\end{center}
\end{figure}

The relevant QNMs for this example are purely decaying, with the relaxation of $J$ and $Q$ governed at late times by the $(+)$-mode, which for the specific $T$ reached here is has $\omega/\sqrt{\rho} \simeq -0.44i$. We demonstrate the agreement of the relaxation of $J,Q$ with this QNM in figure \ref{ringdown2}.

\begin{figure}[h!]
\begin{center}
\vspace{1em}
\includegraphics[width=1\textwidth]{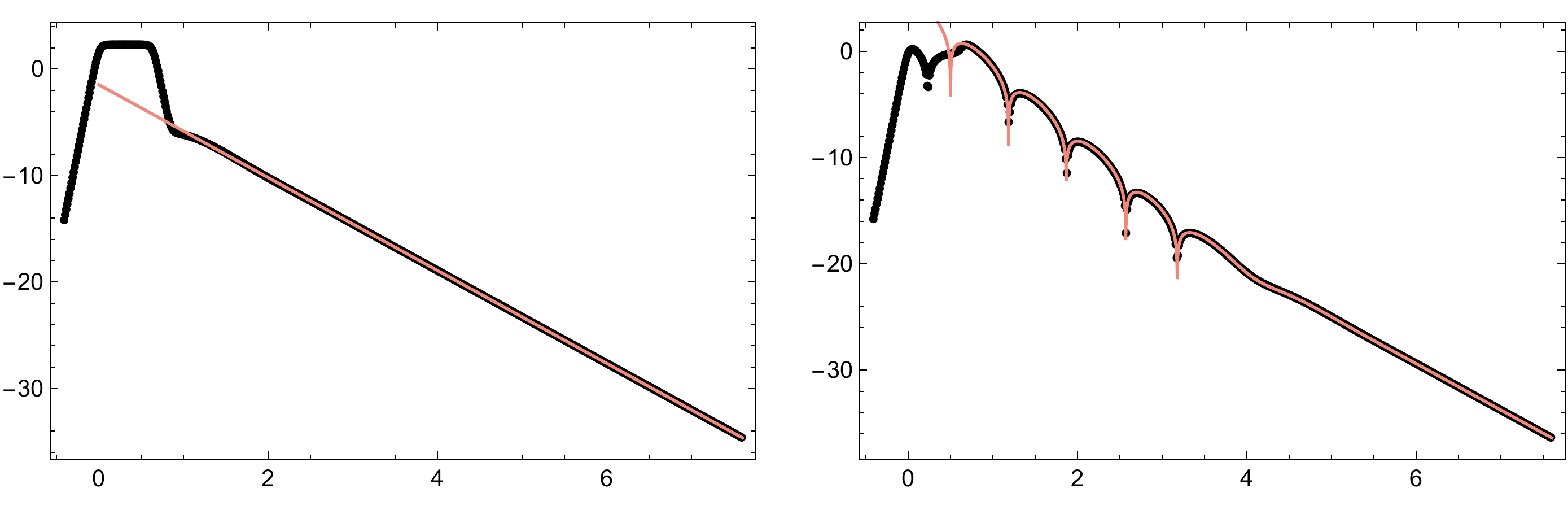}
\begin{picture}(0.1,0.1)(0,0)
\put(-180,163){\makebox(0,0){$\log|J/\rho|$}}
\put(50,163){\makebox(0,0){$\log|Q/\rho^{3/2}|$}}
\put(-108,75){\makebox(0,0){$(+)$}}
\put(165,50){\makebox(0,0){$(+)$}}
\put(85,135){\makebox(0,0){$(-)$}}
\put(-108,15){\makebox(0,0){$\sqrt{\rho}t$}}
\put(116,15){\makebox(0,0){$\sqrt{\rho}t$}}
\put(-108,15){\makebox(0,0){$\sqrt{\rho}t$}}
\end{picture}
\caption{Exponential decay of $J$ (\emph{left panel}) and ringdown of $Q$ (\emph{right panel}) in an incoherent regime at $k/\sqrt{\rho} = 10$ following a quench to $T/\sqrt{\rho} \simeq 0.695$. The QNM portrait near this temperature is shown in figure \ref{QNMs}. The longest lived QNM belongs to the $(+)$-sector and is purely decaying, dominating $J$ and the very late time behaviour of $Q$. Crucially, at large $k$ a shorter-lived contribution from the $(-)$-sector governs $Q$ for some time, which is both oscillating and decaying, making an incoherent regime easily recognisable. The red curves show a fit to these QNMs. \label{ringdown10}}
\end{center}
\end{figure}

For our second example, in figure \ref{ringdown10} we show $J$ and $Q$ in the case $k/\sqrt{\rho} = 10$, for a quench which reaches equilibrium at $T/\sqrt{\rho} \simeq 0.695$. At this temperature the longest lived $(+)$-sector mode has $\omega/\sqrt{\rho} \simeq - 4.36i$, and the longest lived $(-)$-sector mode has $\omega/\sqrt{\rho} \simeq 4.58 - 6.73 i$. For $Q$ we show a fit of a linear combination of these two QNMs. For $J$ we fit only the $(+)$-sector mode. As we previously argued, it is expected that the $(-)$-sector make a dominant contribution to $Q$ despite being much shorter lived, since in the limit $k\to\infty$, $Q$ is sensitive only to this sector. Indeed in figure \ref{ringdown10} we see the oscillatory-decaying behaviour of $Q$ (or \emph{ringdown}, appropriating the dual black hole terminology), but not of $J$, as expected. Eventually the contribution of the longer lived $(+)$-sector mode appears in $Q$ due to $1/k$ effects; we have verified that increasing $k$ extends the time over which the $(-)$-sector mode governs $Q$.

\section{Nonlinear conductivity\label{constE}}
In order to investigate nonlinear thermoelectric response we quench the system away from equilibrium with a step profile, 
\be
E(t) =  \Theta(t)  E_f
\ee
where $\Theta(t)$ is defined in section \ref{hatE}. 

\subsection{Defining an effective temperature\label{Teff}}
In the numerical sections that follow it will be useful to have a notion of effective temperature, $T_E(t)$, during time evolution. 
We define $T_E(t)$ as the temperature of the equilibrium state to which the system eventually settles down if we stop driving the system at time $t$.
This definition has the benefit of carrying a clear physical interpretation at any time, even far from equilibrium, and it is precisely the temperature when the system is in a stable equilibrium. 
It also extends beyond the use of just the local energy density, $\epsilon(t)$, accounting for the possibility of other time dependent charges. 
In the presence of an instability, determining $T_E(t)$ clearly can become an involved task, and whether or not it would be a useful quantity in such cases remains to be seen.

In all the examples given here, no instabilities are seen and the late time equilibrium solutions are known, given in section \ref{model}. Moreover at fixed $k$ and $\rho$ (which are both constants for the evolution) equilibrium can be labelled by the energy density, $\epsilon$, thus we can easily compute $T_E(t)$ by computing the temperature of the equilibrium state with energy $\epsilon(t)$.

\subsection{Linear regime\label{steadylinear}}
\begin{figure}[h!]
\begin{center}
\includegraphics[width=0.9\textwidth]{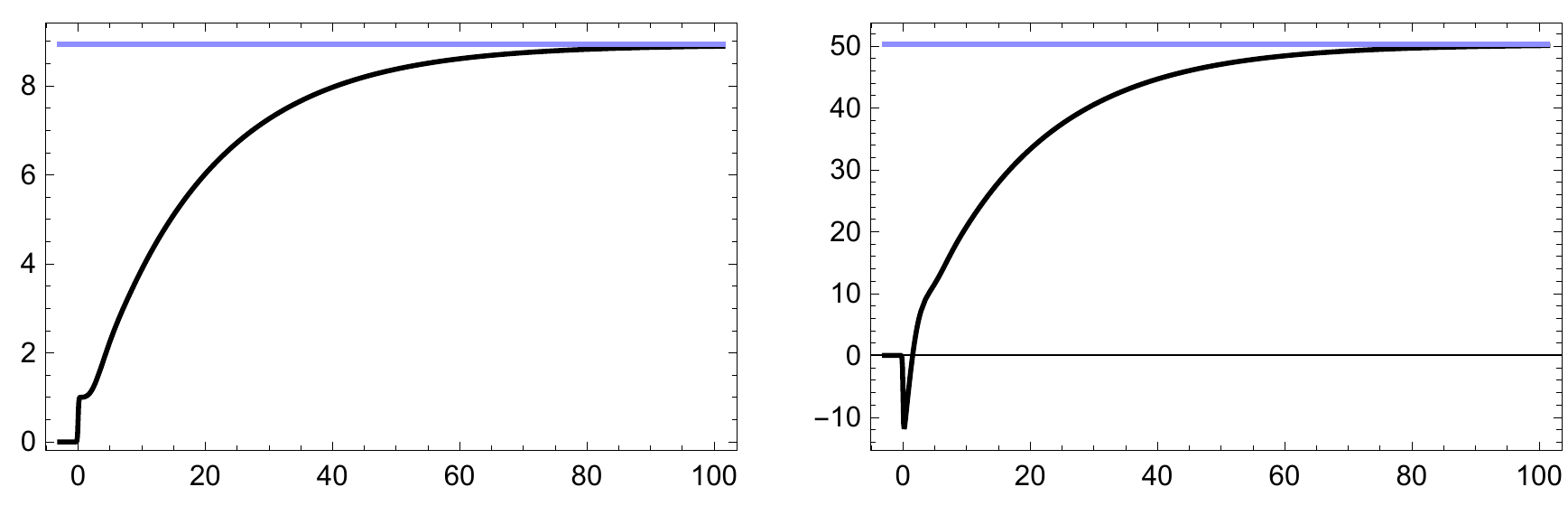}
\begin{picture}(0.1,0.1)(0,0)
\put(-300,-5){\makebox(0,0){$\sqrt{\rho} t$}}
\put(-95,-5){\makebox(0,0){$\sqrt{\rho} t$}}
\put(-401,93){\makebox(0,0){$\frac{J(t)}{E_f}$}}
\put(-197,93){\makebox(0,0){$\frac{Q(t)}{E_fT}$}}
\end{picture}
\caption{Electric and heat currents in the linear response regime. Here we begin at $T_i/\sqrt{\rho} = 1/10$ with $k/\sqrt{\rho} =1/2$ and $E_f/\rho = 10^{-3}$. The blue lines show the expected steady state values given DC linear response electric and thermoelectric conductivities \eqref{linearcond}. The timescale of the approach is set by the longest lived QNM of the equilibrium black brane. \label{linfig}}
\end{center}
\end{figure}
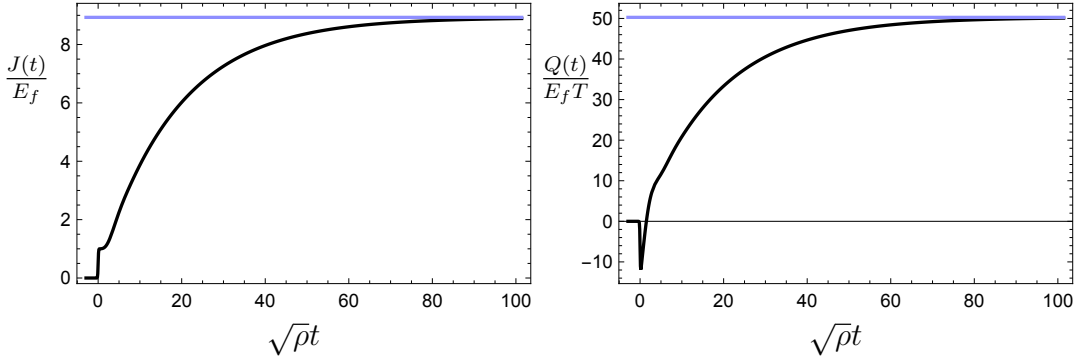

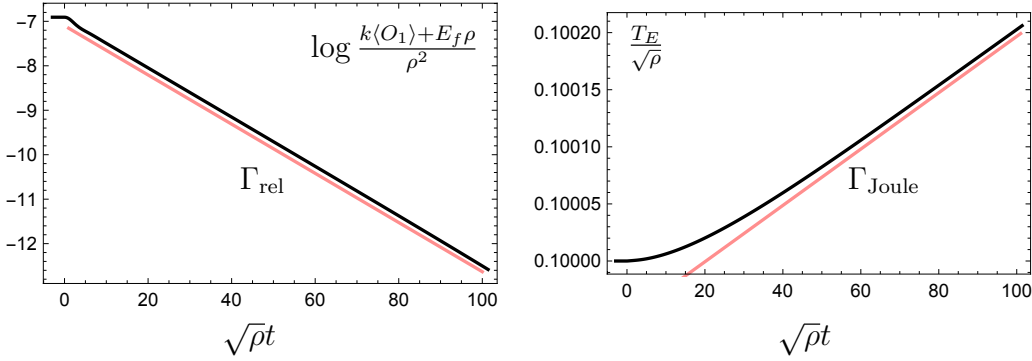
\begin{figure}[h!]
\begin{center}
\includegraphics[width=0.9\textwidth]{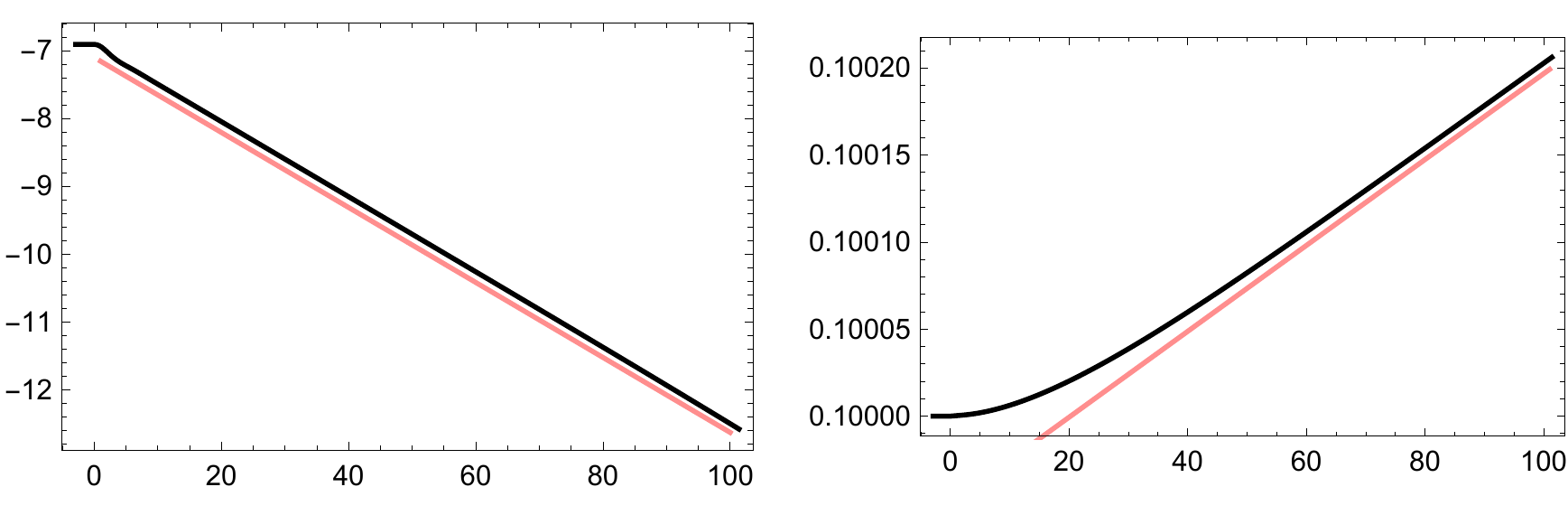}
\begin{picture}(0.1,0.1)(0,0)
\put(-245,105){\makebox(0,0){$\log \frac{k \left<O_1\right> + E_f \rho}{\rho^2}$}}
\put(-150,105){\makebox(0,0){$\frac{T_E}{\sqrt{\rho}}$}}
\put(-300,-5){\makebox(0,0){$\sqrt{\rho} t$}}
\put(-90,-5){\makebox(0,0){$\sqrt{\rho} t$}}
\put(-295,55){\makebox(0,0){$\Gamma_\text{rel}$}}
\put(-60,55){\makebox(0,0){$\Gamma_\text{Joule}$}}
\end{picture}
\vspace{1em}
\caption{Approach to the linear steady state for the evolution presented in figure \ref{linfig} in the linear response regime. \emph{Left panel:} Convergence of $\left<O_1\right>$ to its steady state value, $\left<O_1\right> = -E_f\rho/k$. The red line shows the expected slope given by the momentum relaxation rate given by the longest lived QNM at equilibrium. \emph{Right panel:} Time dependence of the effective temperature defined in section \ref{Teff}. The red line shows the rate of effective temperature increase predicted from the equilibrium black brane.\label{linVTfig}}
\end{center}
\end{figure}

Here we take the example of $E_f/\rho = 10^{-3}$ showing $J(t)/E_f$ and $Q(t)/E_f/T$ in figure \ref{linfig} together with the expected values assuming DC linear response, \eqref{linearcond}.  As anticipated the immediate response is for $J$ to track $E$ during the quench, corresponding to the kink from $J/E_f =0$ to $J/E_f=1$ around $t = 0$. Following this, momentum grows until it is balanced by the momentum sink, and the charged, linear response steady state is achieved. In figure \ref{linVTfig} we verify the approach to the steady state is governed by the longest lived vector QNM, by plotting the scalar vev, which in the steady state takes the value  $\left<O_1\right> = -E_f\rho/k$. 

In this example the electric field is small but finite, and so there is a small Joule heating effect, with a rate of temperature increase, $\Gamma_\text{Joule}$. Near equilibrium the Ward Identity \eqref{joule} can be used to give,
\be
\Gamma_\text{Joule} \equiv \frac{1}{T}\frac{\partial T}{\partial t}\bigg|_\rho = \frac{1}{T} \frac{1}{c_\rho}\sigma E^2\label{eqmheat}
\ee
where $c_\rho$ is the specific heat, to be evaluated for the equilibrium black brane. Deviations will occur for strong electric fields, or after significant heating. We show $T_E(t)$ in the right panel of figure \ref{linVTfig}, together with the timescale $\Gamma_\text{Joule}$.

\subsection{Examples with significant Joule heating\label{notsteady}}
For a steady state we may reasonably expect $J$ to depend on $T/\sqrt{\rho}$ -- as indeed it does in linear response -- but we may also expect $J$ to depend nonlinearly on $E_f$.
In the absence of a steady state there is the additional complication of time dependence.
However, if the rate of heating, $\Gamma_\text{Joule}$, is sufficiently low, we may approximate the time dependence of $J$ through its temperature dependence by promoting $T\to T_E(t)$,
\be
J  =  \sigma\left( \frac{k}{\sqrt{\rho}}, \frac{T_E(t)}{\sqrt{\rho}}, \frac{E_f}{\rho}\right) E_f. \label{Japprox}
\ee
Higher order corrections in the heating rate $\Gamma_\text{Joule}$ could also be considered systematically in a derivative expansion, but for now we focus on this leading effect.

We shall demonstrate that the $T_E$ dependence is remarkably simple -- good agreement is achieved by taking the DC linear response result for $\sigma$, which may be written in terms of the temperature at equilibrium, $T$, and promoting $T \to T_E(t)$. In other words, $\sigma$ does not appear to depend on $E_f$ explicitly. Let us refer to $\sigma$ computed in this way as $\sigma_\text{lin}$. A practical short-cut for this is to first eliminate $\mu$ from $\sigma$ \eqref{linearcond} using the thermodynamic relations \eqref{thermoEp} and \eqref{thermoRho} to obtain,
\be
\sigma_\text{lin} = 1 + \frac{\mu(k,\rho,\epsilon)^2}{k^2}.
\ee 
Since $\rho$ is conserved and $k$ is fixed, it is only the energy density, $\epsilon$, which we update during the evolution. Note that the linear response result for $\bar{\alpha}$ \eqref{linearcond} is constant since $\rho$ is conserved.

\begin{figure}[h!]
\begin{center}
\includegraphics[width=0.9\textwidth]{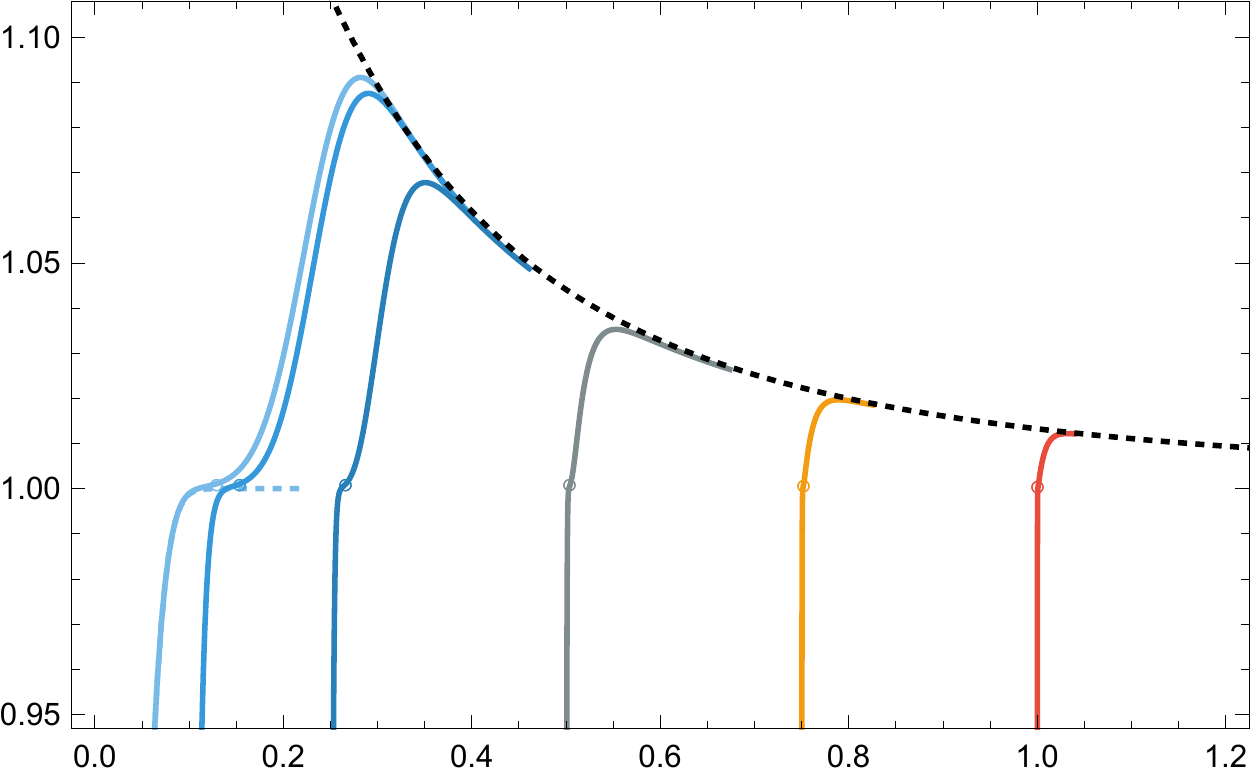}
\begin{picture}(0.1,0.1)(0,0)
\put(-400,195){\makebox(0,0){$\frac{J}{E_f}$}}
\put(-245,225){\makebox(0,0){$\sigma_\text{lin}\left( \frac{k}{\sqrt{\rho}}, \frac{T_E}{\sqrt{\rho}}\right)$}}
\put(-355,35){\makebox(0,0){${\scriptstyle \frac{1}{100}}$}}
\put(-320,35){\makebox(0,0){${\scriptstyle \frac{1}{10}}$}}
\put(-280,35){\makebox(0,0){${\scriptstyle \frac{1}{4}}$}}
\put(-210,35){\makebox(0,0){${\scriptstyle \frac{1}{2}}$}}
\put(-135,35){\makebox(0,0){${\scriptstyle \frac{3}{4}}$}}
\put(-55,35){\makebox(0,0){${\scriptstyle \frac{T_i}{\sqrt{\rho}}\; =\; 1}$}}
\put(-190,-10){\makebox(0,0){$T_E/\sqrt{\rho}$}}
\end{picture}
\vspace{1em}
\caption{Nonlinear electrical current response at $E_f/\rho = 1$, as a function of effective temperature $T_E$ (defined in section \ref{Teff}). Under the applied electric field $T_E$ increases and the system evolves from left to right along the solid curves shown, which represent runs of differing initial temperatures $T_i/\sqrt{\rho} = 10^{-2}, 10^{-1}, 1/4,1/2,3/4,1$ as labelled. Constant $E$ is reached after the open circles shown, at which point $(E - E_f)/E_f \simeq 10^{-5}$. After some time the attractor behaviour is reached, marked by the black dashed line, which is the DC linear response conductivity after the equilibrium temperature is promoted to $T_E$, as described in the text. The blue dashed curve shows the behaviour $J=E$ for the run $T_i/\sqrt{\rho} = 10^{-2}$. Here $k/\sqrt{\rho} = 2$. \label{attractorfig}}
\end{center}
\end{figure}
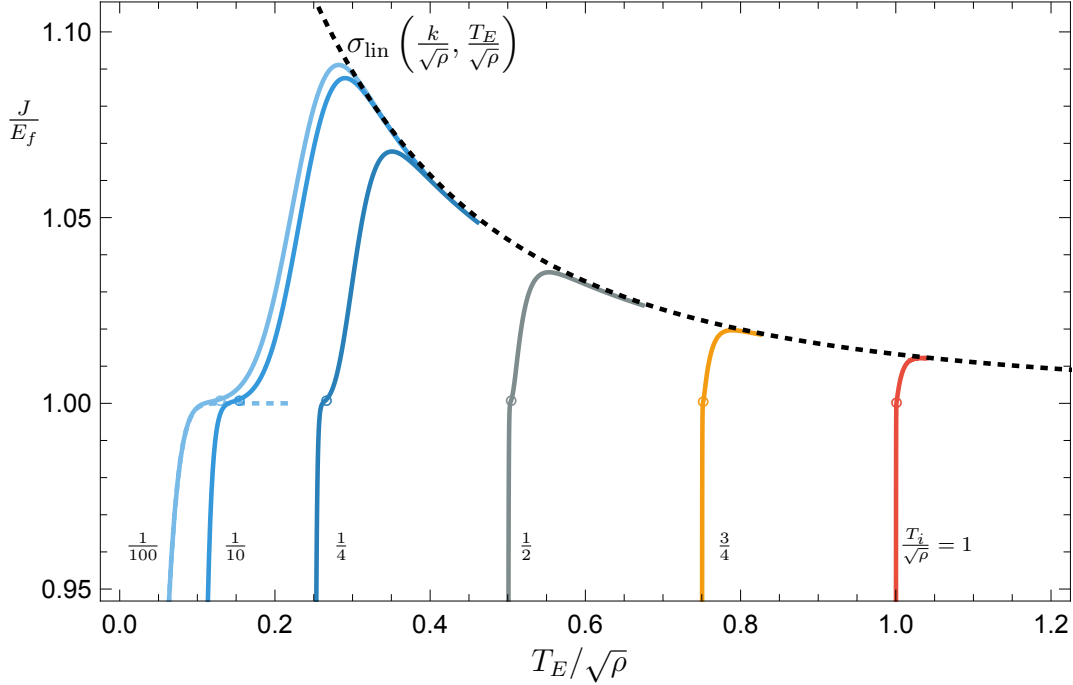

In figure \ref{attractorfig} we show $J/E_f$ as measured during the evolution as a function of $T_E$ at $E_f/\rho = 1$. After some transient period $\sigma$ approaches the attractor governed by the linear response approximation, independently of the initial temperature.  We have verified that in the regime where the linear response result applies, subleading corrections due to higher derivatives of $T_E$ are small.\footnote{Note similar agreement for the $\sigma$ approximation can be observed in example of the top hat electric field profile presented in section \ref{hatE}, corresponding to the dashed blue curve in figure \ref{stepbulk}.} 

We have also verified similar behaviour for the momentum relaxation values, $k/\sqrt{\rho} = 1/2$ and $k/\sqrt{\rho} = 8$. We note that for large $k/\sqrt{\rho}$ one can construct examples where $\sigma$ and $\bar{\alpha}$ are approximately constant in the presence of Joule heating, at non-negligible charge densities. In particular $\sigma$ and $\bar\alpha$ are given by their linear response values. This closely resembles the situation at $\rho=0$, and it would be interesting to try and construct them in perturbation theory about the Vaidya-like solutions of section \ref{neutral}.

The plateau emerging at low $T_E$ in figure \ref{attractorfig} occurs during a period where $E$ is still time dependent, as indicated by the open circles, which indicate the point at which $E(t)$ becomes approximately constant. 
The dashed curve shows the value $J=E$ for the run $T_i/\sqrt{\rho} = 10^{-2}$. The plateau coincides with a period of rapid temperature increase; one possible explanation for it is then that the growth in temperature is much faster than the response time due to the presence of charge and the system appears neutral, hence the instantaneous $J=E$ behaviour just as in section \ref{neutral}. 

\begin{figure}[h!]
\begin{center}
\includegraphics[width=0.9\textwidth]{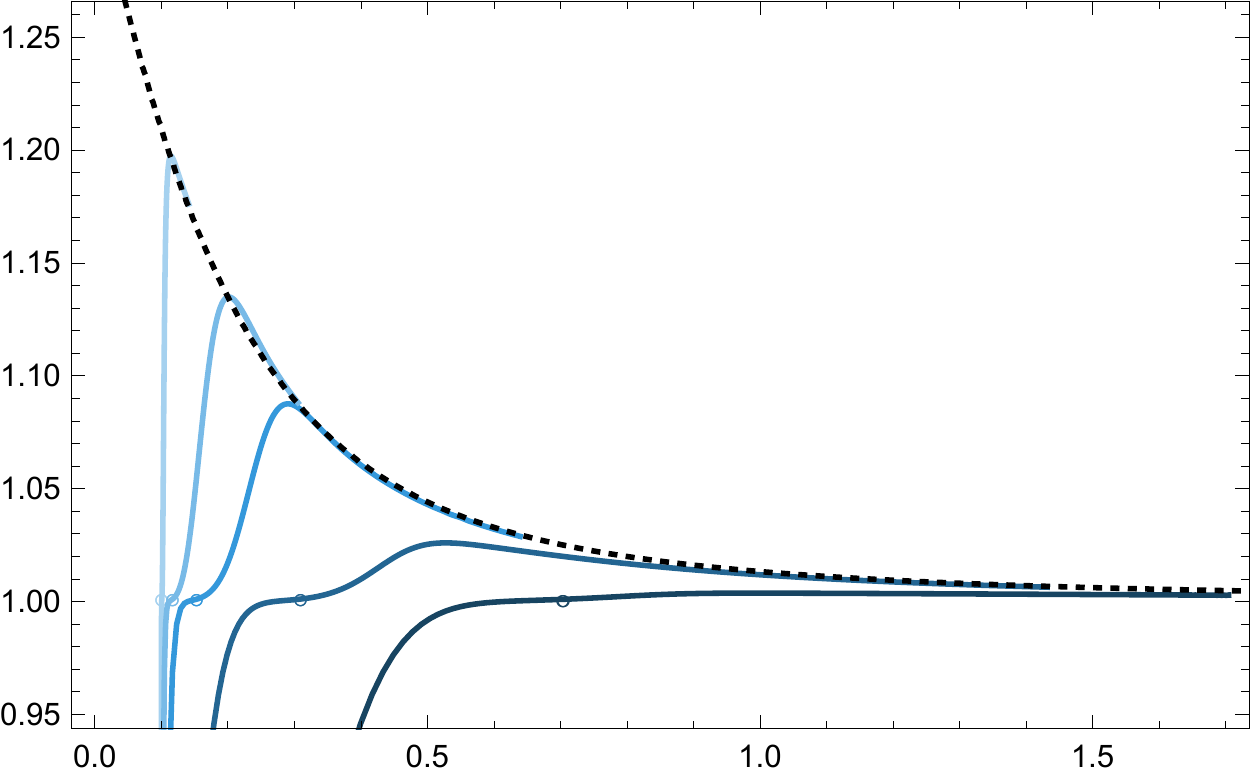}
\begin{picture}(0.1,0.1)(0,0)
\put(-405,175){\makebox(0,0){$\frac{J}{E_f}$}}
\put(-305,225){\makebox(0,0){$\sigma_\text{lin}\left( \frac{k}{\sqrt{\rho}}, \frac{T_E}{\sqrt{\rho}}\right)$}}
\put(-350,135){\makebox(0,0){${\scriptstyle \frac{1}{10}}$}}
\put(-325,105){\makebox(0,0){${\scriptstyle \frac{1}{2}}$}}
\put(-310,85){\makebox(0,0){${\scriptstyle 1}$}}
\put(-290,65){\makebox(0,0){${\scriptstyle 3}$}}
\put(-245,35){\makebox(0,0){${\scriptstyle \frac{E_f}{\rho}\; =\;10}$}}
\put(-190,-10){\makebox(0,0){$T_E/\sqrt{\rho}$}}
\end{picture}
\vspace{1em}
\caption{Nonlinear electrical current response for various $E_f/\rho$, as a function of effective temperature $T_E$ starting from a temperature $T_i/\sqrt{\rho} = 1/10$. Under the applied electric field $T_E$ increases and the system evolves from left to right along the solid curves shown, which represent runs of differing $E_f/\rho = 1/10,1/2,1,3,10$ as labelled. Constant $E$ is reached roughly after the open circles shown, at which point $(E - E_f)/E_f \simeq 10^{-5}$. After some time the attractor behaviour is reached, marked by the black dashed line, which is the DC linear response conductivity after the equilibrium temperature is promoted to $T_E$, as described in the text. The amount of heating before the linear response regime is reached increases with $E_f$, leading to a higher $T_E$ value when they eventually agree. Here $k/\sqrt{\rho} = 2$. \label{Evary}}
\end{center}
\end{figure}

Finally we illustrate the dependence on $E_f$ in figure \ref{Evary}, up to $E_f/\rho = 10$. The linear response curve, $\sigma_\text{lin}$ is eventually reached for each case. $\Gamma_\text{Joule}$ is higher for larger $E_f$, therefore the system is at a higher $T_E$ by the time any transients have settled down and $\sigma_\text{lin}$ is reached. For this reason, whether or not the $\sigma_\text{lin}$ approximation applies for arbitrarily large $E_f/T_E^2$ becomes difficult to assess via a quench. Here we have at least good agreement with $\sigma_\text{lin}$ in cases where $E_f/T_E^2 \sim 1$.

\begin{figure}[h!]
\begin{center}
\includegraphics[width=0.9\textwidth]{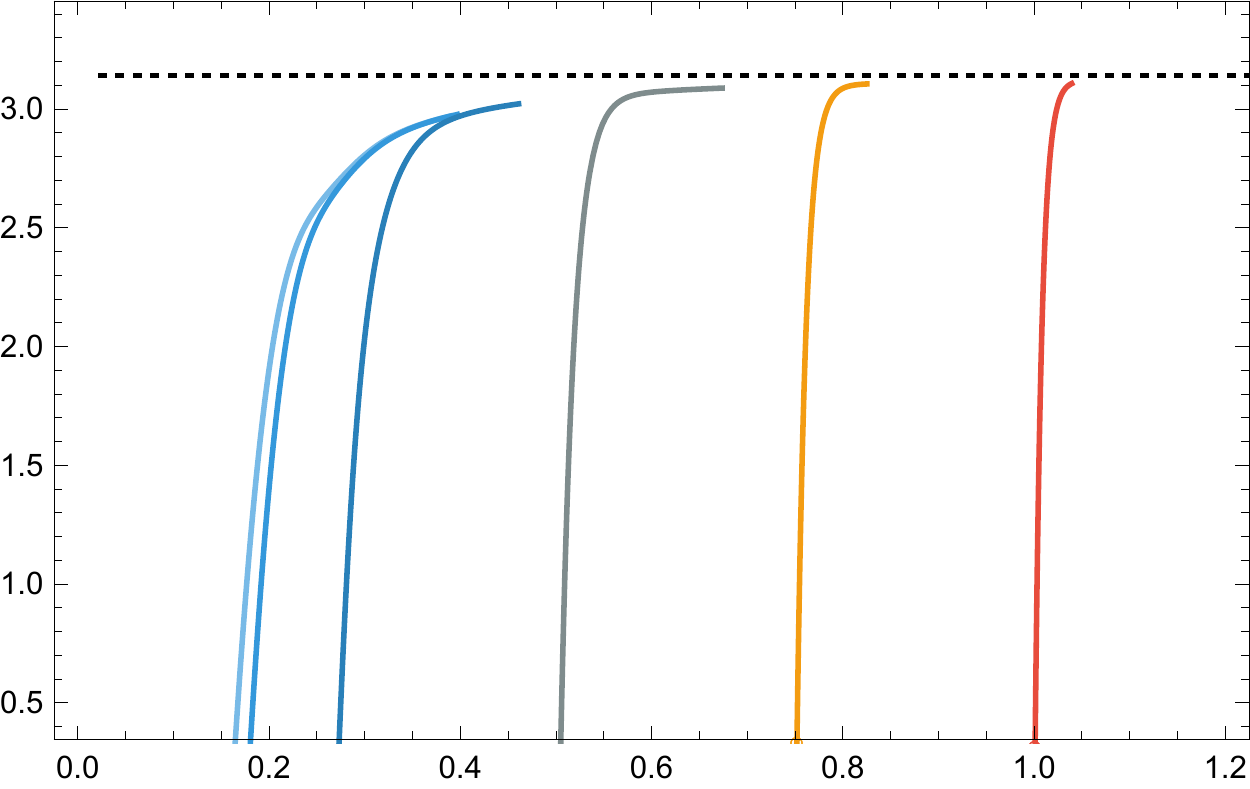}
\begin{picture}(0.1,0.1)(0,0)
\put(-400,195){\makebox(0,0){$\frac{Q}{T_EE_f}$}}
\put(-245,235){\makebox(0,0){$\bar{\alpha}_\text{lin}\left( \frac{k}{\sqrt{\rho}}, \frac{T_E}{\sqrt{\rho}}\right)$}}
\put(-330,35){\makebox(0,0){${\scriptstyle \frac{1}{100}}$}}
\put(-305,35){\makebox(0,0){${\scriptstyle \frac{1}{10}}$}}
\put(-280,35){\makebox(0,0){${\scriptstyle \frac{1}{4}}$}}
\put(-210,35){\makebox(0,0){${\scriptstyle \frac{1}{2}}$}}
\put(-135,35){\makebox(0,0){${\scriptstyle \frac{3}{4}}$}}
\put(-55,35){\makebox(0,0){${\scriptstyle \frac{T_i}{\sqrt{\rho}}\; =\; 1}$}}
\put(-190,-10){\makebox(0,0){$T_E/\sqrt{\rho}$}}
\end{picture}
\vspace{1em}
\caption{Nonlinear heat current response at $E_f/\rho = 1$, as a function of effective temperature $T_E$ (defined in section \ref{Teff}). Under the applied electric field $T_E$ increases and the system evolves from left to right along the solid curves shown, which represent runs of differing initial temperatures $T_i/\sqrt{\rho} = 10^{-2}, 10^{-1}, 1/4,1/2,3/4,1$ as labelled. The black dashed line shows the DC linear response thermoelectric conductivity. Here $k/\sqrt{\rho} = 2$. \label{attractorQfig}}
\end{center}
\end{figure}

We note that the linear response result for $\bar{\alpha}$ does not give good agreement over the same timescales indicating $\bar{\alpha}$ has an explicit dependence on $E_f$. This is demonstrated in figure \ref{attractorQfig}, where at low enough $T_i/\sqrt{\rho}$ at fixed $E_f/\rho = 1$ the effective temperature dependence of $\bar{\alpha}$ deviates from the linear response result, $\bar{\alpha}_\text{lin}$.

\section{Final comments\label{comments}}
We have presented the time evolution of a holographic metal under the influence of an applied electric field at finite temperature. At $\rho = 0$ the system responds instantaneously to the applied electric field, encoded by a Vaidya-like geometry. There is Joule heating but no momentum and no heat current. Such solutions may be useful reference points for the construction of steady states, or as we have investigated here the addition of charge, where they govern the initial response of the system.

At $\rho \neq 0$ we analysed the response of the system using a nonlinear, numerical evolution of the bulk Einstein-Maxwell-axion system of equations. 
First we studied finite-time quenches after which which the system is allowed to return to equilibrium. In particular no instabilities were encountered and the system behaved as expected, returning to equilibrium with the approach governed by the vector QNMs associated to momentum relaxation.

As previously noted, the vector QNM spectrum can exhibit qualitatively different behaviour depending on $k$. We showed the imprint of these features on the relaxation of $J$ and $Q$. Most notably in a large $k$ incoherent regime, $Q$ receives parametrically enhanced contributions from a sector of QNMs which oscillate and decay, readily identifiable in the relaxation of $Q$ following an electric field quench. We note that at fixed $k/\sqrt{\rho}$ a transition from exponential decay to oscillatory decay of $Q$ following a quench can also be obtained by varying $T/\sqrt{\rho}$. These oscillatory QNMs generalise those studied at $\rho =0$ where they are important for the thermal conductivity \cite{Davison:2014lua} in the vicinity of a coherent to incoherent transition. It would be interesting to understand whether the off-axis mode contributing to the oscillations in $Q$ in the incoherent regime are in any way generic.

We note that the situation described here resembles that of \cite{Bhaseen:2012gg}, where the exchange of dominance of purely imaginary and off-axis QNMs impacted the order parameter equilibration after a quench in a holographic model of superfluidity.

Next we studied a step quench to a constant electric field. For sufficiently weak, constant electric fields, an approximate steady state is reached described by linear response. Since the electric field was small but finite, Joule heating still occurs but is only relevant over much longer timescales. Going beyond linear response the effect of Joule heating becomes significant, and the energy density grows. Nevertheless after some initial response time, the electrical conductivity is well approximated by the equilibrium DC linear response result once the increasing effective temperature (as defined in section \ref{Teff}) is taken into account, $\sigma_\text{lin}$. 

The agreement of the nonlinear evolution with $\sigma_\text{lin}$ is somewhat surprising; in general we should have expected a dependence on $E_f$. This is not unprecedented however, since at $\rho = 0$ we showed that $\sigma = 1$ at any $E(t)$, just as in the case $\rho = k = 0$. It would be interesting to consider corrections to this approximation by performing an expansion in time derivatives of $T_E$. It would also be interesting to perturbatively add charge to the $\rho=0$ solutions presented in section \ref{neutral}.

The problem considered here is inherently time dependent due to Joule heating. A natural next step is to prevent the energy in the system from growing without bound, so that we can reach a steady state at fixed temperature at late times. This may resemble a set up where the system is coupled to an external heat bath. Indeed this is the interpretation drawn in \cite{Sonner:2012if} for a steady state in the context of the $D3/D5$ probe brane system. A first glance at the right hand side of \eqref{joule} suggests the possibility of preventing energy growth by turning on a time dependent source for an axion. Note that one example of a stationary solution with a time dependent scalar source is given already in section \ref{model}, by moving away from the present laboratory frame. Another possibility is considering an electric field applied to a strip or to a finite region on the boundary, leaving an infinite undriven heat bath.

Finally we note that the model considered here is a electrical conductor with an AdS$_2\times$R$^2$ infrared scaling behaviour at $T=0$. It would be interesting to contrast nonlinear transport for holographic metals and insulators across a variety of infrared behaviours.\footnote{See \cite{DamleAndSachdev, GreenAndSondhi} for discussions of transport near quantum critical points.} 
Another interesting direction includes different spacetime dimensions, where $\sigma$ is not dimensionless and can lead to qualitatively different behaviour \cite{Horowitz:2013mia}. Additionally it may be worthwhile to consider the nonlinear response of other classes of momentum-relaxing black branes, such as those which include a spatially modulated charge density.   

\section*{Acknowledgements}
We are pleased to thank Hans Bantilan, Richard Davison and Julian Sonner for valuable comments and suggestions. We would especially like to thank Tomas Andrade for early collaboration on related issues.
We would like to acknowledge the hospitality of Technion during the workshop ``Numerical methods for asymptotically AdS spaces'' whilst this work was being completed.
BW is supported by European Research Council grant ERC-2014-StG639022-NewNGR.

%%%%%%%%%%%%%%%%%%%%%%%%%%%%%%%%
\appendix
%%%%%%%%%%%%%%%%%%%%%%%%%%%%%%%%

\section{Details of the numerical method\label{numerics}}
In this section we provide details of the characteristic method used to evolve the coupled Einstein-Maxwell-axion system with boundary sources. Our method follows that of \cite{Bhaseen:2012gg}, with some differences due to anisotropies and the continual injection of energy. We refer to \cite{Chesler:2013lia} for a review of related characteristic methods.
A minimal ansatz in which is sufficient for the task at hand is given by
\bea
ds^2 &=& \frac{1}{r^2}\big(-F(v,r) dv^2  - 2 dv dr +  2 e^{B(v,r)} F_x(v,r) dv dx \nonumber\\
&&\qquad\qquad\qquad\qquad+ S(v,r)( e^{2B(v,r)} dx^2+ e^{-2B(v,r)}dy^2)\big) \label{metricansatz}\\
A &=& (E(v) x + a_v(v,r) )dv + a_x(v,r) dx\\
\phi_1 &=& k\, x + \Phi(v,r)\\
\phi_2 &=& k\, y
\eea
where the AdS boundary lies at $r=0$. With this ansatz the equation of motion for $\phi_2$ is solved. The fields $X\in\{S,B, F_x, \Phi, a_x, a_v\}$ each obey equations with principle part
\be
\partial_v\partial_r X - \frac{(F_x^2 + F S_x)}{2 S_x} \partial_r^2 X = 0 \label{evo}
\ee
and 5 additional equations with principle parts
\bea
\partial_r^2 F + \frac{2F_x}{S} \partial_r^2 F_x  &=& 0 \label{evosup}\\
\partial_r^2 S &=& 0 \label{constraint1}\\
\partial_v^2 S &=& 0 \label{constraint2}\\
\partial_r^2 a_x - e^B \frac{S}{F_x} \partial_r^2 a_v   &=& 0 \label{constraint3}\\
\partial_r^2 F_x - F_x \partial_r^2 B  &=& 0. \label{constraint4}
\eea
Note that there is no longer any $x^I$ dependence explicitly in the equations of motion. Given the fields $X$ and $F$ at time $v$, we use Crank-Nicolson to obtain $X$ at time $v+\Delta v$ using \eqref{evo}, getting $F$ from \eqref{evosup}, making sure to average it between times $v$ and $v+\Delta v$ to ensure second order behaviour. The remaining equations are constraints; we ensure that they are solved for our initial data and consequently they are preserved for the evolution, provided we enforce energy conservation \eqref{joule}. As a note of caution, we have found that using a different choice for \eqref{evosup} in order to perform the evolution can lead to a numerical instability, which can seen by monitoring the constraints for sufficiently long times on an equilibrium solution. With the choice made in \eqref{evosup} the scheme is numerically stable and can be evolved for long times, and with good convergence as we show in Appendix \ref{convergenceconvergenceconvergence}. 

We use Chebyschev collocation in the $r$-direction. It is convenient to factor out the leading terms at the AdS boundary,
\bea
S &=& 1+  2\lambda(v)r + \lambda(v)^2 r^2 +h_{s}(v,r)r^2\\
B &=& h_b(v,r)r^2\\
F_x &=& h_{vx}(v,r)r^2\\
F &=& 1+ 2\lambda(v)r  + \frac{1}{2}(2 \lambda(v)^2 - 4 \lambda'(v)-k^2)r^2 + h_{vv}(v,r)r^2\\
\Phi &=& h(v,r)r^2\\
a_v &=& h_v(v,r)\\
a_x &=& h_x(v,r)
\eea
and work with $Y\in\{h_{vv},h_{vx},h_{s},h_{b},h, h_v, h_x\}$, which satisfy $Y=0$ at the AdS boundary, imposed as a Dirichlet boundary condition. The function $\lambda(t)$ is a gauge parameter, arising from the invariance of the form of the ansatz \eqref{metricansatz} under $r \to r/(1+\lambda(t) r)$. For the majority of evolutions in this paper we use $\lambda$ to adjust the radial coordinate so that the apparent horizon during evolution sits at $r=1$. The only exception is the example shown in figures \ref{stepbulk} and \ref{ringdown2}, where we fix $\lambda = 0$. For initial data we use the analytical black brane solutions presented in \eqref{model}. Correspondingly we arrange our equilibrium initial data such that its horizon is situated at $r=1$, in practise this is achieved by adjusting $\rho$ for some fixed initial temperature, $T_i/\sqrt{\rho}$, and momentum relaxation parameter $k/\sqrt{\rho}$. This will allow us to perform large injections of energy without the need for excision. We do not impose any boundary conditions at $r=1$.

The undetermined data near the boundary associated to the holographic stress tensor and scalar vev is accessible via one radial derivative of the functions $Y$. In particular, after performing holographic renormalisation (see for example \cite{Bianchi:2001kw} and in the context of the axion model, \cite{Andrade:2013gsa})
\bea
\left<T_{tt}\right> &=& -2 h_{vv}^{(1)}\\
\left<T_{tx}\right> = \left<T_{xt}\right>  &=& 3 h_{vx}^{(1)}\\
\left<T_{xx}\right> &=& - h_{vv}^{(1)} + 6 h_{b}^{(1)}\\
\left<T_{yy}\right> &=& - h_{vv}^{(1)} - 6 h_{b}^{(1)}\\
\left<O_x\right> &=& 3 h^{(1)}\\
\left<J_t\right> &=& h_v^{(1)}\\
\left<J_x\right> &=& E(t)+h_x^{(1)}
\eea
with other entries zero. In addition these satisfy \eqref{wardT} and \eqref{wardJ} courtesy of first order equations amongst the $Y^{(1)} \equiv \partial_r Y|_{\text{bdy}}$, as well as $\left<T_\mu^{~\mu}\right> = 0$.

\section{Convergence testing\label{convergenceconvergenceconvergence}}
In this section we present convergence tests over the full dynamical range of one of the evolutions presented. A successful convergence test will show second order convergence with $\Delta v$ (for Crank-Nicolson) and exponential convergence with $N$, the number of Chebyschev points in the $r$-direction. Note that due to the exponential convergence with $N$, the residual becomes dominated by $\Delta v^2$-errors even for relatively small values of $N$, and inevitably it becomes impractical to reduce $\Delta v$ in order to overcome this. 

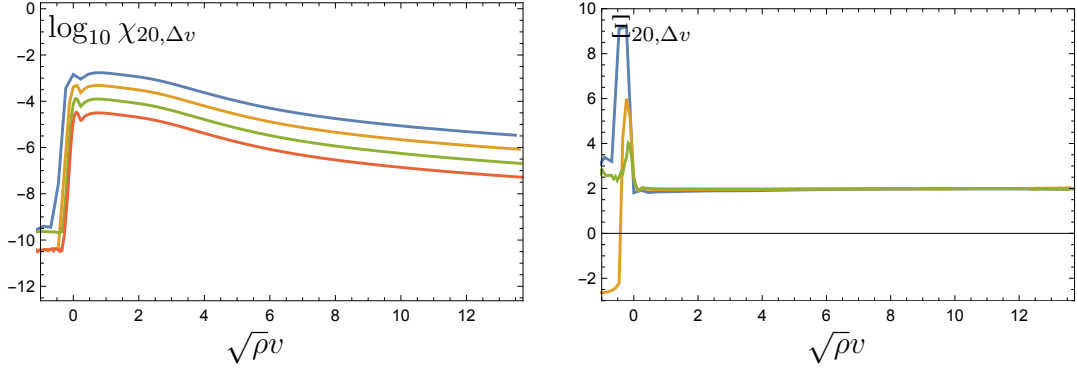
\begin{figure}[h!]
\begin{center}
\includegraphics[width=0.95\textwidth]{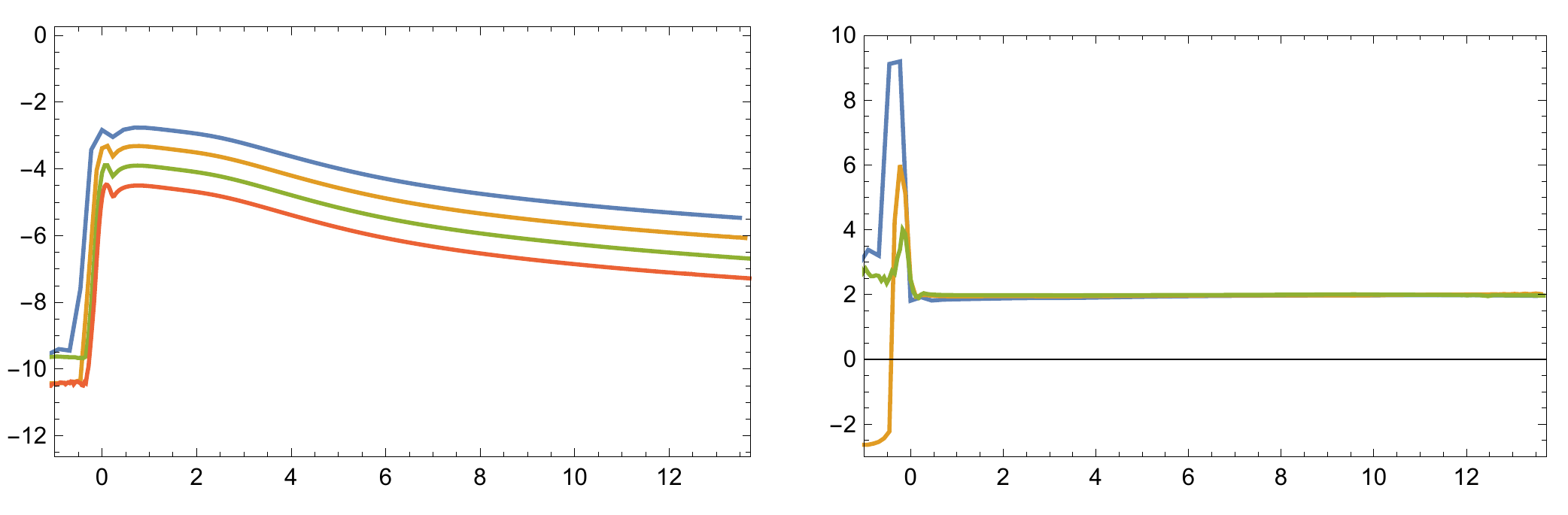}\\
\begin{picture}(0.1,0.1)(0,0)
\put(-160,135){\makebox(0,0){$\log_{10}\chi_{20,\Delta v}$}}
\put(40,135){\makebox(0,0){$\Xi_{20,\Delta v}$}}
\put(-110,15){\makebox(0,0){$\sqrt{\rho}v$}}
\put(110,15){\makebox(0,0){$\sqrt{\rho}v$}}
\end{picture}
\caption{\emph{Left panel:} log of the norm of the constraint equations \eqref{constraint1}, \eqref{constraint2}, \eqref{constraint3}, \eqref{constraint4} as described in the text, for several values of $\sqrt{\rho} \Delta v \simeq 0.915\times(2^{-2},2^{-3},2^{-4},2^{-5})$, from top to bottom. \emph{Right panel:} The quantity $\Xi_{20,\Delta v}$ characterising the convergence rate for refined temporal resolutions; the value $2$ should be approached for the second order scheme used here. Here the colouring corresponds to the $\Delta v$ label as used in the left panel. \label{consplots}}
\end{center}
\end{figure}

We show convergence towards zero for each of the constraints \eqref{constraint1}, \eqref{constraint2}, \eqref{constraint3}, \eqref{constraint4}. At each time $v$ we compute $\chi_{N,\Delta v}$, defined as the $L_\infty$-norm of the vector of constraints, and then take the $L_2$-norm over the spatial grid. Following this we compute, 
\be
\Xi_{N,\Delta v} = \frac{1}{\log 2}\log \frac{\chi_{N,\Delta v}}{\chi_{N,\frac{1}{2}\Delta v}}
\ee
as a function of time. For second order convergence in time, this quantity should approach the value $2$. 
We focus on a single example, $E_f/\rho= 1$, $k/\sqrt{\rho} = 2$, $T_i/\sqrt{\rho} =1/10$. For convergence in $\Delta v$ we show $\chi_{20,\Delta v}$ for a variety of $\Delta v$, and their corresponding $\Xi_{20,\Delta v}$ in figure \ref{consplots}.  

\bibliography{elec}{}

\providecommand{\href}[2]{#2}\begingroup\raggedright\begin{thebibliography}{10}

\bibitem{Horowitz:2012ky}
G.~T. Horowitz, J.~E. Santos and D.~Tong, {\it {Optical Conductivity with
  Holographic Lattices}},  {\em JHEP} {\bf 07} (2012) 168
  [\href{http://arXiv.org/abs/1204.0519}{{\tt 1204.0519}}].
%%CITATION = ARXIV:1204.0519;%%

\bibitem{Horowitz:2012gs}
G.~T. Horowitz, J.~E. Santos and D.~Tong, {\it {Further Evidence for
  Lattice-Induced Scaling}},  {\em JHEP} {\bf 11} (2012) 102
  [\href{http://arXiv.org/abs/1209.1098}{{\tt 1209.1098}}].
%%CITATION = ARXIV:1209.1098;%%

\bibitem{Donos:2012js}
A.~Donos and S.~A. Hartnoll, {\it {Interaction-driven localization in
  holography}},  {\em Nature Phys.} {\bf 9} (2013) 649--655
  [\href{http://arXiv.org/abs/1212.2998}{{\tt 1212.2998}}].
%%CITATION = ARXIV:1212.2998;%%

\bibitem{Horowitz:2013jaa}
G.~T. Horowitz and J.~E. Santos, {\it {General Relativity and the Cuprates}},
  {\em JHEP} {\bf 06} (2013) 087 [\href{http://arXiv.org/abs/1302.6586}{{\tt
  1302.6586}}].
%%CITATION = ARXIV:1302.6586;%%

\bibitem{Ling:2013nxa}
Y.~Ling, C.~Niu, J.-P. Wu and Z.-Y. Xian, {\it {Holographic Lattice in
  Einstein-Maxwell-Dilaton Gravity}},  {\em JHEP} {\bf 11} (2013) 006
  [\href{http://arXiv.org/abs/1309.4580}{{\tt 1309.4580}}].
%%CITATION = ARXIV:1309.4580;%%

\bibitem{Chesler:2013qla}
P.~Chesler, A.~Lucas and S.~Sachdev, {\it {Conformal field theories in a
  periodic potential: results from holography and field theory}},  {\em Phys.
  Rev.} {\bf D89} (2014), no.~2 026005
  [\href{http://arXiv.org/abs/1308.0329}{{\tt 1308.0329}}].
%%CITATION = ARXIV:1308.0329;%%

\bibitem{Balasubramanian:2013yqa}
K.~Balasubramanian and C.~P. Herzog, {\it {Losing Forward Momentum
  Holographically}},  {\em Class. Quant. Grav.} {\bf 31} (2014) 125010
  [\href{http://arXiv.org/abs/1312.4953}{{\tt 1312.4953}}].
%%CITATION = ARXIV:1312.4953;%%

\bibitem{Donos:2014cya}
A.~Donos and J.~P. Gauntlett, {\it {Thermoelectric DC conductivities from black
  hole horizons}},  {\em JHEP} {\bf 11} (2014) 081
  [\href{http://arXiv.org/abs/1406.4742}{{\tt 1406.4742}}].
%%CITATION = ARXIV:1406.4742;%%

\bibitem{Donos:2014oha}
A.~Donos, B.~Goutéraux and E.~Kiritsis, {\it {Holographic Metals and
  Insulators with Helical Symmetry}},  {\em JHEP} {\bf 09} (2014) 038
  [\href{http://arXiv.org/abs/1406.6351}{{\tt 1406.6351}}].
%%CITATION = ARXIV:1406.6351;%%

\bibitem{Donos:2014yya}
A.~Donos and J.~P. Gauntlett, {\it {The thermoelectric properties of
  inhomogeneous holographic lattices}},  {\em JHEP} {\bf 01} (2015) 035
  [\href{http://arXiv.org/abs/1409.6875}{{\tt 1409.6875}}].
%%CITATION = ARXIV:1409.6875;%%

\bibitem{Rangamani:2015hka}
M.~Rangamani, M.~Rozali and D.~Smyth, {\it {Spatial Modulation and
  Conductivities in Effective Holographic Theories}},  {\em JHEP} {\bf 07}
  (2015) 024 [\href{http://arXiv.org/abs/1505.05171}{{\tt 1505.05171}}].
%%CITATION = ARXIV:1505.05171;%%

\bibitem{Davison:2013txa}
R.~A. Davison, K.~Schalm and J.~Zaanen, {\it {Holographic duality and the
  resistivity of strange metals}},  {\em Phys. Rev.} {\bf B89} (2014), no.~24
  245116 [\href{http://arXiv.org/abs/1311.2451}{{\tt 1311.2451}}].
%%CITATION = ARXIV:1311.2451;%%

\bibitem{Lucas:2014zea}
A.~Lucas, S.~Sachdev and K.~Schalm, {\it {Scale-invariant
  hyperscaling-violating holographic theories and the resistivity of strange
  metals with random-field disorder}},  {\em Phys. Rev.} {\bf D89} (2014),
  no.~6 066018 [\href{http://arXiv.org/abs/1401.7993}{{\tt 1401.7993}}].
%%CITATION = ARXIV:1401.7993;%%

\bibitem{Hartnoll:2014cua}
S.~A. Hartnoll and J.~E. Santos, {\it {Disordered horizons: Holography of
  randomly disordered fixed points}},  {\em Phys. Rev. Lett.} {\bf 112} (2014)
  231601 [\href{http://arXiv.org/abs/1402.0872}{{\tt 1402.0872}}].
%%CITATION = ARXIV:1402.0872;%%

\bibitem{Lucas:2015vna}
A.~Lucas, {\it {Conductivity of a strange metal: from holography to memory
  functions}},  {\em JHEP} {\bf 03} (2015) 071
  [\href{http://arXiv.org/abs/1501.05656}{{\tt 1501.05656}}].
%%CITATION = ARXIV:1501.05656;%%

\bibitem{O'Keeffe:2015awa}
D.~K. O'Keeffe and A.~W. Peet, {\it {Perturbatively charged holographic
  disorder}},  {\em Phys. Rev.} {\bf D92} (2015), no.~4 046004
  [\href{http://arXiv.org/abs/1504.03288}{{\tt 1504.03288}}].
%%CITATION = ARXIV:1504.03288;%%

\bibitem{Hartnoll:2015faa}
S.~A. Hartnoll, D.~M. Ramirez and J.~E. Santos, {\it {Emergent scale invariance
  of disordered horizons}},  {\em JHEP} {\bf 09} (2015) 160
  [\href{http://arXiv.org/abs/1504.03324}{{\tt 1504.03324}}].
%%CITATION = ARXIV:1504.03324;%%

\bibitem{Donos:2013eha}
A.~Donos and J.~P. Gauntlett, {\it {Holographic Q-lattices}},  {\em JHEP} {\bf
  04} (2014) 040 [\href{http://arXiv.org/abs/1311.3292}{{\tt 1311.3292}}].
%%CITATION = ARXIV:1311.3292;%%

\bibitem{Andrade:2013gsa}
T.~Andrade and B.~Withers, {\it {A simple holographic model of momentum
  relaxation}},  {\em JHEP} {\bf 05} (2014) 101
  [\href{http://arXiv.org/abs/1311.5157}{{\tt 1311.5157}}].
%%CITATION = ARXIV:1311.5157;%%

\bibitem{Donos:2014uba}
A.~Donos and J.~P. Gauntlett, {\it {Novel metals and insulators from
  holography}},  {\em JHEP} {\bf 06} (2014) 007
  [\href{http://arXiv.org/abs/1401.5077}{{\tt 1401.5077}}].
%%CITATION = ARXIV:1401.5077;%%

\bibitem{Taylor:2014tka}
M.~Taylor and W.~Woodhead, {\it {Inhomogeneity simplified}},  {\em Eur. Phys.
  J.} {\bf C74} (2014), no.~12 3176 [\href{http://arXiv.org/abs/1406.4870}{{\tt
  1406.4870}}].
%%CITATION = ARXIV:1406.4870;%%

\bibitem{Andrade:2016tbr}
T.~Andrade, {\it {A simple model of momentum relaxation in Lifshitz
  holography}},  \href{http://arXiv.org/abs/1602.00556}{{\tt 1602.00556}}.
%%CITATION = ARXIV:1602.00556;%%

\bibitem{Iqbal:2008by}
N.~Iqbal and H.~Liu, {\it {Universality of the hydrodynamic limit in AdS/CFT
  and the membrane paradigm}},  {\em Phys. Rev.} {\bf D79} (2009) 025023
  [\href{http://arXiv.org/abs/0809.3808}{{\tt 0809.3808}}].
%%CITATION = ARXIV:0809.3808;%%

\bibitem{Donos:2015gia}
A.~Donos and J.~P. Gauntlett, {\it {Navier-Stokes Equations on Black Hole
  Horizons and DC Thermoelectric Conductivity}},  {\em Phys. Rev.} {\bf D92}
  (2015), no.~12 121901 [\href{http://arXiv.org/abs/1506.01360}{{\tt
  1506.01360}}].
%%CITATION = ARXIV:1506.01360;%%

\bibitem{Banks:2015wha}
E.~Banks, A.~Donos and J.~P. Gauntlett, {\it {Thermoelectric DC conductivities
  and Stokes flows on black hole horizons}},  {\em JHEP} {\bf 10} (2015) 103
  [\href{http://arXiv.org/abs/1507.00234}{{\tt 1507.00234}}].
%%CITATION = ARXIV:1507.00234;%%

\bibitem{Davison:2013jba}
R.~A. Davison, {\it {Momentum relaxation in holographic massive gravity}},
  {\em Phys. Rev.} {\bf D88} (2013) 086003
  [\href{http://arXiv.org/abs/1306.5792}{{\tt 1306.5792}}].
%%CITATION = ARXIV:1306.5792;%%

\bibitem{Blake:2015epa}
M.~Blake, {\it {Momentum relaxation from the fluid/gravity correspondence}},
  {\em JHEP} {\bf 09} (2015) 010 [\href{http://arXiv.org/abs/1505.06992}{{\tt
  1505.06992}}].
%%CITATION = ARXIV:1505.06992;%%

\bibitem{Davison:2014lua}
R.~A. Davison and B.~Goutéraux, {\it {Momentum dissipation and effective
  theories of coherent and incoherent transport}},  {\em JHEP} {\bf 01} (2015)
  039 [\href{http://arXiv.org/abs/1411.1062}{{\tt 1411.1062}}].
%%CITATION = ARXIV:1411.1062;%%

\bibitem{Davison:2015bea}
R.~A. Davison and B.~Goutéraux, {\it {Dissecting holographic conductivities}},
   {\em JHEP} {\bf 09} (2015) 090 [\href{http://arXiv.org/abs/1505.05092}{{\tt
  1505.05092}}].
%%CITATION = ARXIV:1505.05092;%%

\bibitem{Andrade:2015hpa}
T.~Andrade, S.~A. Gentle and B.~Withers, {\it {Drude in D major}},  {\em JHEP}
  {\bf 06} (2016) 134 [\href{http://arXiv.org/abs/1512.06263}{{\tt
  1512.06263}}].
%%CITATION = ARXIV:1512.06263;%%

\bibitem{Hashimoto:2014yza}
K.~Hashimoto, S.~Kinoshita, K.~Murata and T.~Oka, {\it {Electric Field Quench
  in AdS/CFT}},  {\em JHEP} {\bf 09} (2014) 126
  [\href{http://arXiv.org/abs/1407.0798}{{\tt 1407.0798}}].
%%CITATION = ARXIV:1407.0798;%%

\bibitem{Ali-Akbari:2015gba}
S.~Amiri-Sharifi, H.~R. Sepangi and M.~Ali-Akbari, {\it {Electric Field Quench,
  Equilibration and Universal Behavior}},  {\em Phys. Rev.} {\bf D91} (2015)
  126007 [\href{http://arXiv.org/abs/1504.03559}{{\tt 1504.03559}}].
%%CITATION = ARXIV:1504.03559;%%

\bibitem{Ali-Akbari:2015hoa}
H.~Ebrahim, S.~Heshmatian and M.~Ali-Akbari, {\it {Thermal quench at finite 't
  Hooft coupling}},  {\em Nucl. Phys.} {\bf B904} (2016) 527--537
  [\href{http://arXiv.org/abs/1510.07974}{{\tt 1510.07974}}].
%%CITATION = ARXIV:1510.07974;%%

\bibitem{Karch:2007pd}
A.~Karch and A.~O'Bannon, {\it {Metallic AdS/CFT}},  {\em JHEP} {\bf 09} (2007)
  024 [\href{http://arXiv.org/abs/0705.3870}{{\tt 0705.3870}}].
%%CITATION = ARXIV:0705.3870;%%

\bibitem{Albash:2007bq}
T.~Albash, V.~G. Filev, C.~V. Johnson and A.~Kundu, {\it {Quarks in an external
  electric field in finite temperature large N gauge theory}},  {\em JHEP} {\bf
  08} (2008) 092 [\href{http://arXiv.org/abs/0709.1554}{{\tt 0709.1554}}].
%%CITATION = ARXIV:0709.1554;%%

\bibitem{Karch:2010kt}
A.~Karch and S.~L. Sondhi, {\it {Non-linear, Finite Frequency Quantum Critical
  Transport from AdS/CFT}},  {\em JHEP} {\bf 01} (2011) 149
  [\href{http://arXiv.org/abs/1008.4134}{{\tt 1008.4134}}].
%%CITATION = ARXIV:1008.4134;%%

\bibitem{Sonner:2012if}
J.~Sonner and A.~G. Green, {\it {Hawking Radiation and Non-equilibrium Quantum
  Critical Current Noise}},  {\em Phys. Rev. Lett.} {\bf 109} (2012) 091601
  [\href{http://arXiv.org/abs/1203.4908}{{\tt 1203.4908}}].
%%CITATION = ARXIV:1203.4908;%%

\bibitem{Berridge:2013yba}
A.~M. Berridge and A.~G. Green, {\it {Nonequilibrium conductivity at quantum
  critical points}},  {\em Phys. Rev.} {\bf B88} (2013), no.~22 220512
  [\href{http://arXiv.org/abs/1312.4432}{{\tt 1312.4432}}].
%%CITATION = ARXIV:1312.4432;%%

\bibitem{Baggioli:2016oju}
M.~Baggioli and O.~Pujolas, {\it {On Effective Holographic Mott Insulators}},
  \href{http://arXiv.org/abs/1604.08915}{{\tt 1604.08915}}.
%%CITATION = ARXIV:1604.08915;%%

\bibitem{Kundu:2013eba}
A.~Kundu and S.~Kundu, {\it {Steady-state Physics, Effective Temperature
  Dynamics in Holography}},  {\em Phys. Rev.} {\bf D91} (2015), no.~4 046004
  [\href{http://arXiv.org/abs/1307.6607}{{\tt 1307.6607}}].
%%CITATION = ARXIV:1307.6607;%%

\bibitem{Kundu:2015qda}
A.~Kundu, {\it {Effective Temperature in Steady-state Dynamics from
  Holography}},  {\em JHEP} {\bf 09} (2015) 042
  [\href{http://arXiv.org/abs/1507.00818}{{\tt 1507.00818}}].
%%CITATION = ARXIV:1507.00818;%%

\bibitem{Horowitz:2013mia}
G.~T. Horowitz, N.~Iqbal and J.~E. Santos, {\it {Simple holographic model of
  nonlinear conductivity}},  {\em Phys. Rev.} {\bf D88} (2013), no.~12 126002
  [\href{http://arXiv.org/abs/1309.5088}{{\tt 1309.5088}}].
%%CITATION = ARXIV:1309.5088;%%

\bibitem{Rangamani:2015sha}
M.~Rangamani, M.~Rozali and A.~Wong, {\it {Driven Holographic CFTs}},  {\em
  JHEP} {\bf 04} (2015) 093 [\href{http://arXiv.org/abs/1502.05726}{{\tt
  1502.05726}}].
%%CITATION = ARXIV:1502.05726;%%

\bibitem{Bardoux:2012aw}
Y.~Bardoux, M.~M. Caldarelli and C.~Charmousis, {\it {Shaping black holes with
  free fields}},  {\em JHEP} {\bf 05} (2012) 054
  [\href{http://arXiv.org/abs/1202.4458}{{\tt 1202.4458}}].
%%CITATION = ARXIV:1202.4458;%%

\bibitem{Hartnoll:2016tri}
S.~A. Hartnoll, D.~M. Ramirez and J.~E. Santos, {\it {Entropy production,
  viscosity bounds and bumpy black holes}},  {\em JHEP} {\bf 03} (2016) 170
  [\href{http://arXiv.org/abs/1601.02757}{{\tt 1601.02757}}].
%%CITATION = ARXIV:1601.02757;%%

\bibitem{Bhaseen:2012gg}
M.~J. Bhaseen, J.~P. Gauntlett, B.~D. Simons, J.~Sonner and T.~Wiseman, {\it
  {Holographic Superfluids and the Dynamics of Symmetry Breaking}},  {\em Phys.
  Rev. Lett.} {\bf 110} (2013), no.~1 015301
  [\href{http://arXiv.org/abs/1207.4194}{{\tt 1207.4194}}].
%%CITATION = ARXIV:1207.4194;%%

\bibitem{DamleAndSachdev}
K.~{Damle} and S.~{Sachdev}, {\it {Nonzero-temperature transport near quantum
  critical points}},  {\em Phys. Rev.} {\bf B56} (Oct., 1997) 8714--8733
  [\href{http://arXiv.org/abs/cond-mat/9705206}{{\tt cond-mat/9705206}}].

\bibitem{GreenAndSondhi}
A.~G. Green and S.~L. Sondhi, {\it Nonlinear quantum critical transport and the
  schwinger mechanism for a superfluid-mott-insulator transition of bosons},
  {\em Phys. Rev. Lett.} {\bf 95} (Dec, 2005) 267001.

\bibitem{Chesler:2013lia}
P.~M. Chesler and L.~G. Yaffe, {\it {Numerical solution of gravitational
  dynamics in asymptotically anti-de Sitter spacetimes}},  {\em JHEP} {\bf 07}
  (2014) 086 [\href{http://arXiv.org/abs/1309.1439}{{\tt 1309.1439}}].
%%CITATION = ARXIV:1309.1439;%%

\bibitem{Bianchi:2001kw}
M.~Bianchi, D.~Z. Freedman and K.~Skenderis, {\it {Holographic
  renormalization}},  {\em Nucl. Phys.} {\bf B631} (2002) 159--194
  [\href{http://arXiv.org/abs/hep-th/0112119}{{\tt hep-th/0112119}}].
%%CITATION = HEP-TH/0112119;%%

\end{thebibliography}\endgroup
\bibliographystyle{JHEP-2}

\end{document}